\documentclass{ieeeaccess}

\usepackage{authblk}
\usepackage{placeins}
\usepackage{graphicx}
\usepackage[colorlinks]{hyperref}
\usepackage{textcomp}
\usepackage{xspace}
\usepackage{caption}
\usepackage{subcaption}
\usepackage{booktabs}
\usepackage{textcomp}
\usepackage{amsmath}
\usepackage{balance}
\usepackage{multirow}
\usepackage{multicol}

\def\BibTeX{{\rm B\kern-.05em{\sc i\kern-.025em b}\kern-.08em
T\kern-.1667em\lower.7ex\hbox{E}\kern-.125emX}}

\newcommand{\ModelName}{BlazeNeo\xspace}
\newcommand{\DHA}{BlazeNeo-DHA\xspace}
\newcommand{\DIA}{BlazeNeo-DIA\xspace}
\newcommand{\IDA}{BlazeNeo-IDA\xspace}
\newcommand{\LSC}{BlazeNeo-LSC\xspace}
\newcommand{\DatasetName}{NeoPolyp\xspace}
\newcommand{\CleanDatasetName}{NeoPolyp-Clean\xspace}

\begin{document}

\title{\ModelName{}: Blazing fast polyp segmentation and neoplasm detection}
\doi{10.1109/ACCESS.2021.DOI}
%
% \titlerunning{Blazing fast polyp segmentation and neoplasm detection}

% \author{
%   Nguyen Sy An\inst{1}
%   \and Phan Ngoc Lan\inst{1}
%   \and Dao Viet Hang\inst{2,3}
%   \and Dao Van Long\inst{2,3}
%   \and Tran Quang Trung\inst{4}
%   \and Nguyen Thi Thuy\inst{5}
%   \and Dinh Viet Sang\inst{1}
% }

% \authorrunning{Nguyen et al.}

% \institute{
%   Hanoi University of Science and Technology, Hanoi, Vietnam
%   \and Hanoi Medical University, Hanoi, Vietnam
%   \and The Institute of Gastroenterology and Hepatology, Hanoi, Vietnam
%   \and University of Medicine and Pharmacy, Hue University, Hue, Vietnam
%   \and Faculty of Information Technology, Vietnam National University of Agriculture, Hanoi, Vietnam
% }

% \maketitle

\author{
    \uppercase{Nguyen S. An}\authorrefmark{1},
    \uppercase{Phan N. Lan}\authorrefmark{1},
    \uppercase{Dao V. Hang}\authorrefmark{2,3},
    \uppercase{Dao V. Long}\authorrefmark{2,3},
    \uppercase{Tran Q. Trung}\authorrefmark{4},
    \uppercase{Nguyen T. Thuy}\authorrefmark{5},
    \uppercase{Dinh V. Sang}\authorrefmark{1}
}

\address[1]{Hanoi University of Science and Technology, Hanoi, Vietnam}
\address[2]{Hanoi Medical University, Hanoi, Vietnam}
\address[3]{Institute of Gastroenterology and Hepatology, Hanoi, Vietnam}
\address[4]{University of Medicine and Pharmacy, Hue University, Hue, Vietnam}
\address[5]{Faculty of Information Technology, Vietnam National University of Agriculture, Hanoi, Vietnam}

\corresp{Corresponding author: Dinh V. Sang (e-mail: sangdv@soict.hust.edu.vn).}

\begin{abstract}
    In recent years, computer-aided automatic polyp segmentation and neoplasm detection have been an emerging topic in medical image analysis, providing valuable support to colonoscopy procedures. Attentions have been paid to improving the accuracy of polyp detection and segmentation. However, not much focus has been given to latency and throughput for performing these tasks on dedicated devices, which can be crucial for practical applications. This paper introduces a novel deep neural network architecture called \ModelName, for the task of polyp segmentation and neoplasm detection with an emphasis on compactness and speed while maintaining high accuracy. The model leverages the highly efficient HarDNet backbone alongside lightweight Receptive Field Blocks for computational efficiency, and an auxiliary training mechanism to take full advantage of the training data for the segmentation quality. Our experiments on a challenging dataset show that \ModelName achieves improvements in latency and model size while maintaining comparable accuracy against state-of-the-art methods. When deploying on the Jetson AGX Xavier edge device in INT8 precision, our \ModelName achieves over 155 fps while yielding the best accuracy among all compared methods.
\end{abstract}

\titlepgskip=-15pt

\maketitle

\section{Introduction}

% \[ F1 = 2\frac{p \cdot r}{p+r}\ \ \mathrm{where}\ \ p = \frac{tp}{tp+fp},\ \ r = \frac{tp}{tp+fn} \]

Colorectal polyps, especially adenomas with high-grade dysplasia, carry high risks of progressing into colorectal cancer (CRC) \cite{gschwantler2002high}, which claims over 640,000 lives each year \cite{bernal2017comparative}. There are available procedures to screen and detect high-risk polyps in an early stage, increasing chances of successful treatment. Polyp detection and removal in colonoscopy are the most effective method to prevent colorectal cancer \cite{issa2017colorectal}.

At the same time, practical factors such as overloading healthcare systems, low-quality endoscopy equipment, or personnel's lack of experience \cite{armin2015visibility,lee2008adequate} can severely limit the effectiveness of colonoscopy. A review by Leufkens et al. \cite{leufkens2012factors} pointed out that $20-47\%$ of polyps might have been missed during colonoscopies. Several types of image-enhanced endoscopies and accessories have been proposed to alleviate these, yet they can be prohibitively expensive for practical applications, especially in poor medical facilities. On the other hand, computer-aided systems for colonoscopy have shown a lot of promise and have attracted many researchers in recent years. Several works have achieved very good performance on benchmark datasets \cite{fan2020pranet,huang2021hardnet,tang2019cu}.

Polyp segmentation is a subset of medical image analysis that has gained much attention recently. Traditional machine learning methods for solving the problem are mostly based on hand-crafted features \cite{iwahori2013automatic,silva2014toward} to extract image information such as color, shape, and textures. Since polyps have very high intra-class diversity and low inter-class variation, such approaches are often limited in representing and detecting polyps. Deep neural networks, and especially U-Net \cite{ronneberger2015u} have been the state-of-the-art methods for polyp segmentation in the last few years. These networks can learn highly abstract and complex features, allowing them to achieve good performances. At the same time, deep neural networks also come with a complexity trade-off, as models can be very large (up to several hundred million parameters) and cause high latency during inference.

Most of the existing research in endoscopy image analysis has focused on the polyp segmentation problem, in which lesion regions or polyps are segmented from the background pixels. Those works attempted to improve the learning models to provide accurate segmentation of polyps. However, the segmentation of polyps only does not provide information about the type of polyps, i.e., benign (non-cancerous) or malignancy (presence of cancerous cells).
In our seminal work \cite{lan2021neounet}, we have defined the problem of Polyp Segmentation and Neoplasm Detection (PSND), aiming at fine-grained segmentation of polyps. This can be considered as an extension of the polyp segmentation problem, providing richer semantic information for the segmented regions. Particularly, besides classifying image pixels as polyp or background, the proposed formulation further identifies each polyp pixel as non-neoplastic, neoplastic, or undefined. In general, non-neoplastic polyps are typically benign, while neoplastic polyps have a risk of developing cancer.
Our newly developed UNet-based neural network architecture, NeoUNet, has obtained state-of-the-art performance in solving this problem in terms of accuracy. However, attention has not yet been paid to the model size and speed, which is challenging for practical deployment.

In this paper, we further improve on our previous works on Polyp Segmentation and Neoplasm Detection with the proposal of \ModelName{}, a novel deep neural network architecture with an efficient learning mechanism. Our main contributions are:
\begin{itemize}
    %\item To describe the design of \ModelName{}, a deep neural network architecture for polyp segmentation and neoplasm detection;
    %\item To present experimental results for \ModelName on the newly collected \DatasetName dataset with comparisons to existing models;
    %\item To present latency and throughput metrics for \ModelName and baseline models on embedded hardware, similar to real-life deployments of polyp segmentation and neoplasm detection.
    \item A new deep neural network architecture called \ModelName{} for polyp segmentation and neoplasm detection, aiming at reducing the model size and therefore improving inference speed;
    \item An auxiliary training strategy to fully exploit training data for maintaining high accuracy while reducing model size;
    \item Extensive experiments on the newly collected \DatasetName dataset and comparisons to existing models. Moreover, we measure model latency and throughput on dedicated hardware in a setting similar to real-life deployments of polyp segmentation and neoplasm detection.
\end{itemize}

The rest of the paper is organized as follows. We provide a brief review of related works in Section \ref{sec:related}. Section \ref{sec:problem} briefly describes the polyp segmentation and neoplasm detection problem as formulated in \cite{lan2021neounet}. The \ModelName architecture is presented in Section \ref{sec:propose}. Section \ref{sec:experiment} showcases our experimental studies. Finally, we conclude the paper and highlight future works in Section \ref{sec:conclude}.

\section{Related work}
\label{sec:related}
In recent years, many computer vision tasks have seen massive improvements through the advancements of convolutional neural networks (CNNs). AlexNet \cite{krizhevsky2012imagenet} and VGG \cite{simonyan2014very} are among the first successful CNNs for the image classification problem. However, these early models still suffer from degradation when increasing network depth. Many works have attempted to modify the network architectures to improve learning capability aiming at improving network performance. Skip connections, first introduced in ResNet \cite{he2016deep} in 2016, helped alleviate the degradation and smoothed out the loss landscape. ResNeXt \cite{xie2017aggregated} combined the idea of skip connections with a multi-branch design first proposed by the authors of GoogLeNet \cite{szegedy2015going}. More recently, Tan et al. \cite{tan2019efficientnet} employed neural architecture search to produce EfficientNet, a family of neural networks with varying levels of the trade-off between accuracy and latency. Meanwhile, HarDNet \cite{chao2019hardnet} is a model highly focused on optimizing inference latency and memory traffic.

Many CNN architectures have been designed for the semantic image segmentation task, especially in medical images. Among the earliest was the work by Long et al. \cite{long2015fully}, who adopted several well-known architectures using transfer learning. In the same year, U-Net \cite{ronneberger2015u} became a breakthrough model in medical imaging, achieving highly promising results for medical image segmentation. Later works such as UNet++ \cite{zhou2019unet++}, DoubleUNet \cite{jha2020doubleu} and Coupled U-Net \cite{tang2019cu} further improved and alleviated limitations in U-Net. ColonSegNet \cite{jha2021real} was a lightweight encoder-decoder architecture that uses residual connections with squeeze and excitation network as the main component. ColonSegNet achieved a high inference speed but with sacrificing accuracy a lot. DDANet \cite{tomar2020ddanet} was another encoder-decoder design that leverages the strength of residual connection and squeeze-and-excitation modules. DDANet incorporated a single encoder followed by two dual decoders. The first decoder is used for the segmentation mask, while the second one acts as an autoencoder model that reconstructs the grayscale image and helps strengthen the feature representation of the encoder. Attention-UNet \cite{oktay2018attention} proposed attention gates as a filter mechanism for selecting useful salient features. Development on non-UNet models in medical segmentation has also remained active. DeepLabV3 \cite{chen2017rethinking} is a prominent architecture that utilizes atrous convolutions for dense feature extraction. Fan et al. \cite{fan2020pranet} enhanced an FCN-like model with parallel partial decoder and reverse attention to form PraNet, a network that achieved state-of-the-art performance on many benchmark datasets. HarDNet-MSEG \cite{huang2021hardnet} employed an encoder-decoder structure with HarDNet as the encoder backbone, achieving good performance on the Kvasir-SEG dataset and very high inference speed. Meanwhile, TransFuse \cite{zhang2021transfuse} combined a CNN with the Transformer architecture using a fusion module called BiFusion.

Many deep learning methods are also specially designed for the polyp segmentation and detection problem. Qadir et al. \cite{qadir2019framework} proposed a framework that incorporates a CNN architecture for labeling segmentation masks for polyps. The framework allows doctors to receive pre-annotations from a model trained in a semi-supervised manner. In \cite{shin2018automatic}, the authors proposed a model with an Inception-ResNet backbone combined with several post-learning methods to enhance polyp detection accuracy. Shin et al. \cite{shin2018abnormal} used conditional adversarial networks to generate abnormal samples for training polyp detection models. Liu et al. \cite{liu2019colonic} applied different CNN backbones including InceptionV3 \cite{szegedy2015going}, ResNet50 \cite{he2016deep} and VGG16 \cite{simonyan2014very} to the SSD  framework, whose accuracy was much higher than other one-stage object detectors and comparable to the Faster-RCNN two-stage one. In \cite{li2021colonoscopy}, Li et al. compared the performance of eight state-of-the-art deep learning object detectors and demonstrated promising results in colonoscopy.

% TO DO: Bo sung cac nghien cuu neoplasm classification
While there have been few works in the past several years concerning polyp neoplasm detection (such as that of Ribeiro et al. \cite{ribeiro2016exploring}), it was only very recently that neoplasm detection has been incorporated into polyp segmentation. Phan et al. \cite{lan2021neounet} formally described the problem (denoted as PSND) and proposed NeoUNet, a UNet-based architecture that established the baseline for PSND. While NeoUNet strikes a balance between accuracy and inference speed, typical deployment scenarios for polyp segmentation systems require relatively low latency to ensure smooth operation. Moreover, such scenarios would have the neural network run on lightweight systems embedded within the colonoscopy device. Therefore, this paper focuses on building a network architecture optimized for speed and small in size.

% Due to the heterogeneous quality of different endoscopic systems, the lack of public datasets for polyp characterization is a challenge. Among the most similar research to ours is that of Ribeiro et al. \cite{ribeiro2016exploring}, who presented several approaches - including CNNs and hand-crafted features - for polyp classification. The authors extracted a dataset of 100 polyp images from endoscopy videos, each containing exactly one polyp. Several CNN models were tested on this dataset, including VGG, AlexNet, GoogLeNet, etc\dots\ A primary drawback for this approach is that classification has to be done after detection or segmentation. In other words, the problem is approached in two stages. While combining the polyp detection modules with classification is possible, this method can be inefficient and cumbersome, especially for systems with real-time requirements or running on embedded devices. In this paper, we present an end-to-end model for both the segmentation and classification of polyps to overcome these limitations.
% To our knowledge, this problem has not been addressed in any previous work.

\begin{figure*}[ht!]
    \begin{center}
        \includegraphics[width=0.9\textwidth]{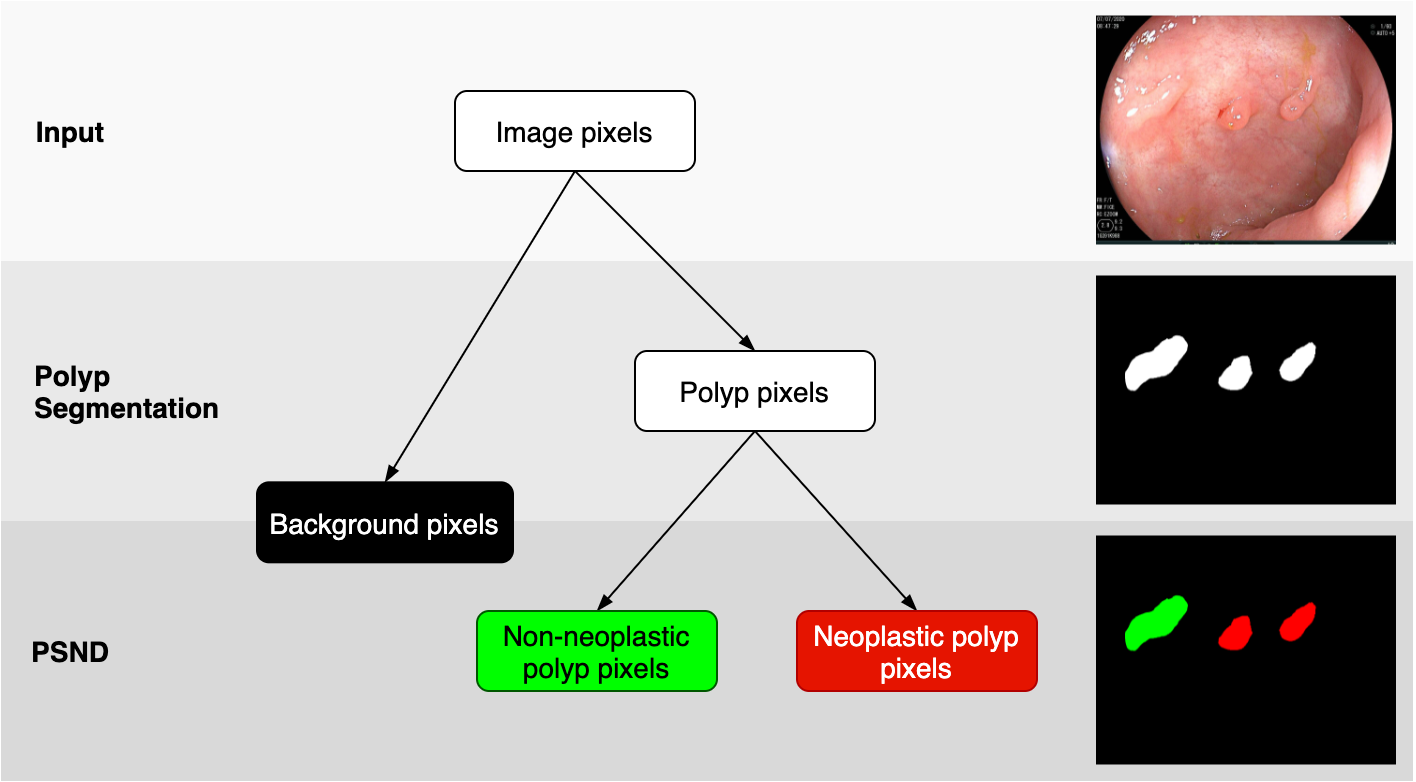}
    \end{center}
    \caption{PSND is an extension of polyp segmentation, which further discriminates whether a polyp is neoplastic or non-neoplastic. For a given input image shown on the top right corner, expected outputs for polyp segmentation and PSND are depicted on the middle and the bottom images on the right, respectively. The black color denotes background pixels, white color denotes polyp regions; green and red colors denote non-neoplastic and neoplastic polyps, respectively.}
    \label{fig:psnd}
\end{figure*}

\section{Polyp Segmentation and Neoplasm Detection}
\label{sec:problem}

Polyp Segmentation and Neoplasm Detection (PSND) has been formulated as a type of fine-grained polyp segmentation problem \cite{lan2021neounet}. Besides segmentation of polyps, this formulation further classifies a polyp pixel into two classes: non-neoplastic or neoplastic. In medical image analysis, non-neoplastic polyps are considered benign, while neoplastic polyps may further progress with a higher risk of cancer. During a colonoscopy, the operator must decide the types of polyps, neoplasm or non-neoplastic, to consider an optimal management strategy, i.e., removal or resection in endoscopy procedure or biopsy and operation. Figures \ref{fig:psnd} depicts the PSND problem as a semantic extension of the polyp segmentation problem.

In practice, an ``undefined'' class is labeled if there is not enough information from the endoscopic image to categorize the risk of neoplasm. This undefined subgroup is not a specific class we want to predict since such predictions would not bring any insight for the endoscopist. Therefore, the model only needs to learn to discriminate between neoplastic and non-neoplastic polyps. However, the undefined class still gives supplementary information about polyp regions that can be exploited to help the model gain more representation power in the training phase.

% \begin{figure}[ht!]
% \centering
% \subcaptionbox{Input image}{
% \includegraphics[width=0.14\textwidth]{images/example.jpeg}
% }
% \subcaptionbox{Polyp\\segmentation}{
% \includegraphics[width=0.14\textwidth]{images/example-seg.png}
% }
% \subcaptionbox{PSND}{
% \includegraphics[width=0.14\textwidth]{images/example-label.png}
% }
% \caption{Expected outputs for polyp segmentation and PSND: (a) Input image; (b) Expected output for the polyp segmentation task. Black regions denote background pixels. White regions denote polyp regions; (c) Expected output for the neoplasm segmentation task. Green and red colors denote non-neoplastic and neoplastic polyps, respectively.}
% \label{fig:example}
% \end{figure}

\section{Proposed Method}
\label{sec:propose}

\begin{figure*}[ht!]
    \begin{center}
        \includegraphics[width=\textwidth]{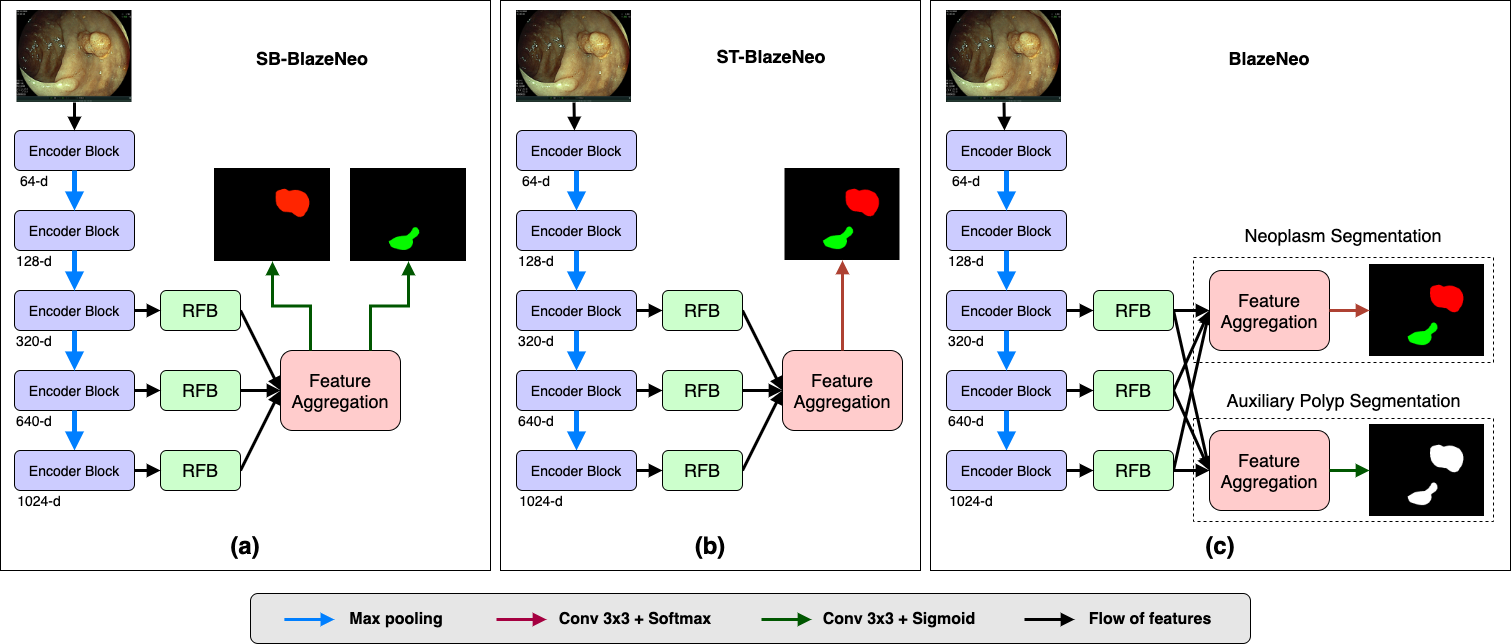}
    \end{center}
    \caption{Proposed architectures of our \ModelName: (a) Single-headed Binary BlazeNeo (SB-BlazeNeo) has one output branch that produces two binary segmentation maps corresponding to neoplastic and non-neoplastic classes; (b) Single-headed Trinary BlazeNeo (ST-BlazeNeo) also has one output branch that directly predicts a trinary segmentation map; (c) Multi-headed BlazeNeo (or BlazeNeo for short) contains two output branches that are responsible for the two tasks: neoplasm segmentation treated as the main task, and polyp segmentation treated as the auxiliary task. Both branches share the same architecture of the feature aggregation module, but they are trained separately without sharing their parameters.}
    \label{fig:model}
\end{figure*}

Figure~\ref{fig:model} depicts three versions of the proposed \ModelName model. The models are developed based on HarDNet-MSEG~\cite{huang2021hardnet}, with several improvements tailored for the PSND problem. First, we use a simplified version of RFB~\cite{liu2018receptive} with smaller kernel sizes and apply different feature aggregation schemes. For the encoder layer, we keep the HarDNet-68~\cite{chao2019hardnet} backbone as in HarDNet-MSEG.

The three \ModelName variants differ in how outputs from the model are generated and processed. Inspired by NeoUNet \cite{lan2021neounet}, Single-headed Binary \ModelName (SB-\ModelName) solves two binary segmentation tasks corresponding to neoplastic and non-neoplastic classes.
The second variant, Single-headed Trinary \ModelName (ST-\ModelName), predicts a trinary map for three classes in the neoplasm segmentation task. Finally, Multi-headed \ModelName (or \ModelName for short) uses two output branches. The main output branch is used for the neoplasm segmentation task, and the auxiliary branch is for the polyp segmentation task. The following subsections describe our design and motivations in detail.

\subsection{Lightweight Encoder: HarDNet}

\begin{center}
    \begin{figure}[th]
        \begin{center}
            \includegraphics[width=0.48\textwidth]{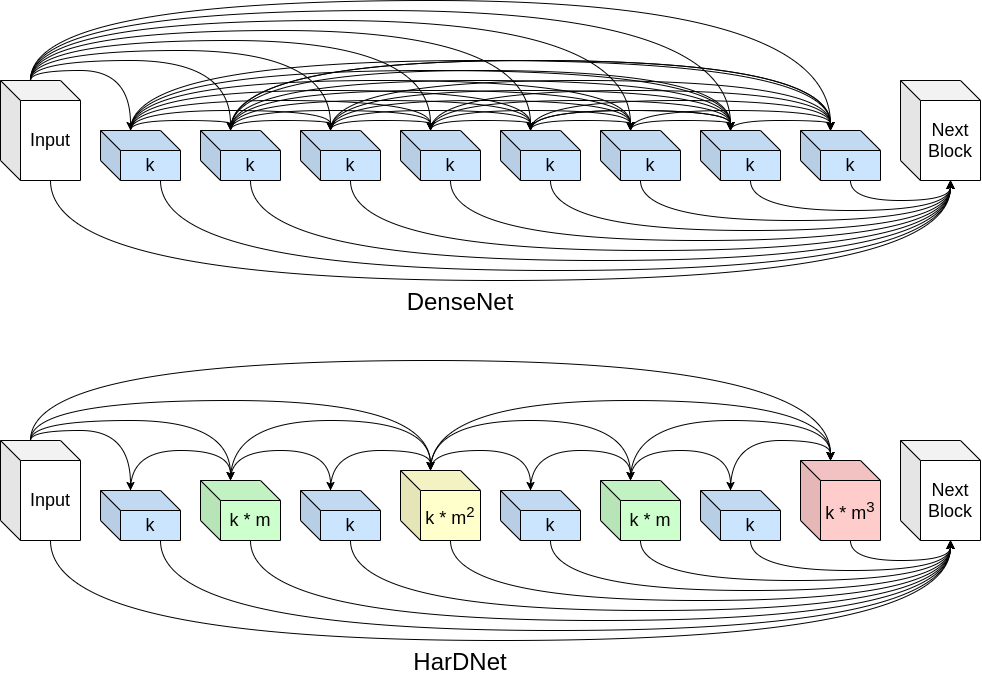}
        \end{center}
        \caption{Illustrations for DenseNet block and Harmonic DenseNet (HarDNet) block. Each of the layers is a $3\times3$ convolution. The value on each layer denotes the number of output channels.}
        \label{fig:hardnet}
    \end{figure}
\end{center}

HarDNet (Harmonic Densely Connected Network)~\cite{chao2019hardnet} is an improvement over DenseNet~\cite{huang2017densely}. The primary goal in HarDNet's design is to lower latency by reducing memory traffic. The authors argued that the connection pattern of Dense Blocks, in which each layer has skip connections toward every proceeding layer in the block, causes ineffective memory access during runtime that severely hinders performance. HarDNet reduces the number of skip connections to form a pattern similar to the harmonic wave function, as well as scaling channel width according to a layer's influence level. We illustrate the difference between the two architectures in Figure~\ref{fig:hardnet}.

Inside a HarDNet block, each layer is indexed from the input layer 0. A layer $l$ receives a skip connection from layer $l-2^n$ if $2^n$ divides $l$ ($n \geq 0$, $l-2^n \geq 0$). Given the initial growth rate $k$ and compression factor $m$, layer $l$'s channel width is equal to $k \times m^x$, where $x = max \{\nu \ |\ l\ \vdots\ 2^{\nu} \}$.

\begin{center}
    \begin{figure}[th]
        \begin{center}
            \includegraphics[width=0.45\textwidth]{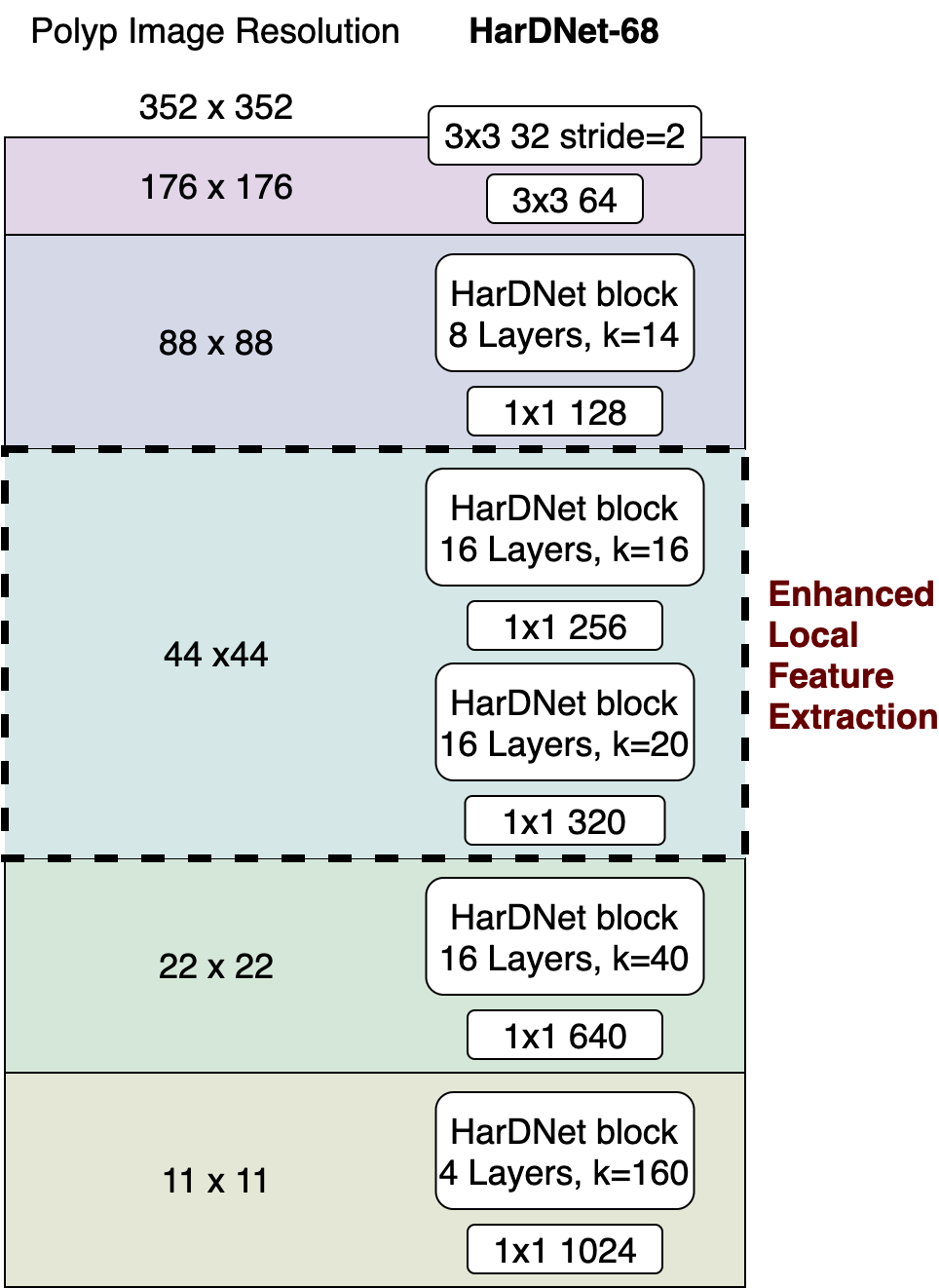}
        \end{center}
        \caption{An overview of HarDNet-68 architecture. Following each HarDNet block is a transitional Conv 1x1 layer.}
        \label{fig:hardnet-68}
    \end{figure}
\end{center}

Chao et al. \cite{chao2019hardnet} further proposed HarDNet-68 for small-object detection problems. While most CNNs focus on stride-16 to enhance classification ability, HarDNet-68 distributes most of the layers on stride-8 to aid small-scale object detection, as shown in Figure~\ref{fig:hardnet-68}. Experiments in~\cite{chao2019hardnet} show that HarDNet-68 is not only 30\% faster than ResNet-50~\cite{he2016deep} but also more accurate than ResNet-101~\cite{he2016deep} when used as backbone for SSD~\cite{liu2016ssd} in the object detection problem.

Our \ModelName also uses HarDNet-68 as the encoder backbone. For an input colonoscopy image $I$ with size  ${h \times w}$, five levels of features ${\{f_{i}, i = 1, 2, 3, 4, 5\}}$ with resolution ${[h/2^{i-1}, w/2^{i-1}]}$ are produced from the encoder. Wu et al. \cite{wu2019cascaded} showed that low-level features (corresponding to $f_1$ and $f_2$) contribute less to performance while being computationally expensive due to their size. Therefore, our \ModelName discards $f_1$ and $f_2$, using only the final three feature maps for the decoder module to accelerate its inference speed.

\subsection{Parallel Partial Decoder}
In all three variants of our \ModelName, the decoder module consists of a Receptive Field Block (RFB) series. Three last feature maps of the encoder are independently passed through the RFB blocks and then fused by the feature aggregation blocks.
% The RFB module works as a transitional unit generating three refined feature maps $\{X_{1}, X_{2}, X_{3}\}$. Then they are aggregated by multiple schemes for both the polyp segmentation task and the neoplasm segmentation task.

\subsubsection{Receptive Field Block}

Polyps can appear in various scales on endoscopic images depending on their actual size, their distance to the colonoscopy camera, or the angle between them and the camera. This is a challenge for CNN architectures, in which the receptive field size is often fixed. Several studies have suggested different mechanisms to create more robust receptive fields, including the Inception block~\cite{szegedy2015going}, ASPP block~\cite{chen2017deeplab}, and Deformable Convolution block~\cite{dai2017deformable}. Conceptually, these proposals are similar in that they all use multiple convolutional branches with different kernel sizes, merging the outputs to form adaptive receptive fields. However, they also have their own limitations. The Inception block samples all the kernels of each branch at their center, ignoring crucial edge details due to small sampling coverage. Meanwhile, the ASPP and Deformable Convolution blocks do not differentiate between different pixel positions, making it difficult for the model to focus on segmentation targets.

The Receptive Field Block (RFB) \cite{liu2018receptive} has been proposed to address these limitations. RFB also uses the multi-branch convolution approach, with improvements inspired by the human visual cortex. In addition, it highlights the relationship between receptive field size and eccentricity, where the distance to the center is proportional to the weight specified or the importance level.

\begin{center}
    \begin{figure}[th]
        \begin{center}
            \includegraphics[scale=.3]{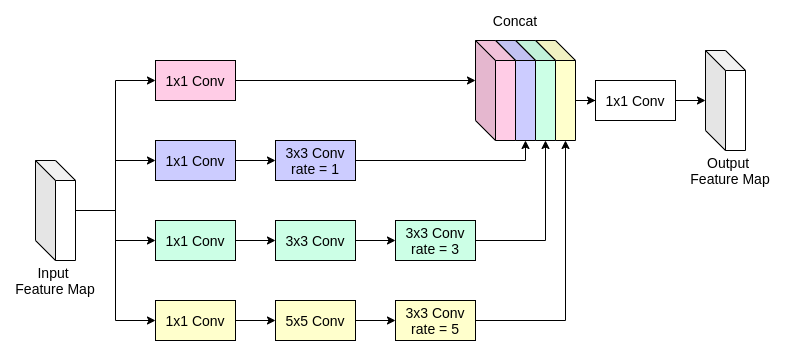}
        \end{center}
        \caption{Structure of the Receptive Field Block (RFB).}
        \label{fig:rfb}
    \end{figure}
\end{center}

The RFB module used in PraNet \cite{fan2020pranet} and HarDNet-MSEG \cite{huang2021hardnet} is modified with a larger kernel size and larger dilation rate. As this version of RFB is more computationally expensive, our \ModelName instead uses a simplified version of RFB (see Figure~\ref{fig:rfb}) as proposed in \cite{liu2018receptive} for faster inference.

\subsubsection{Feature Aggregation}
\begin{center}
    \begin{figure}[th]
        \begin{center}
            \includegraphics[scale=.25]{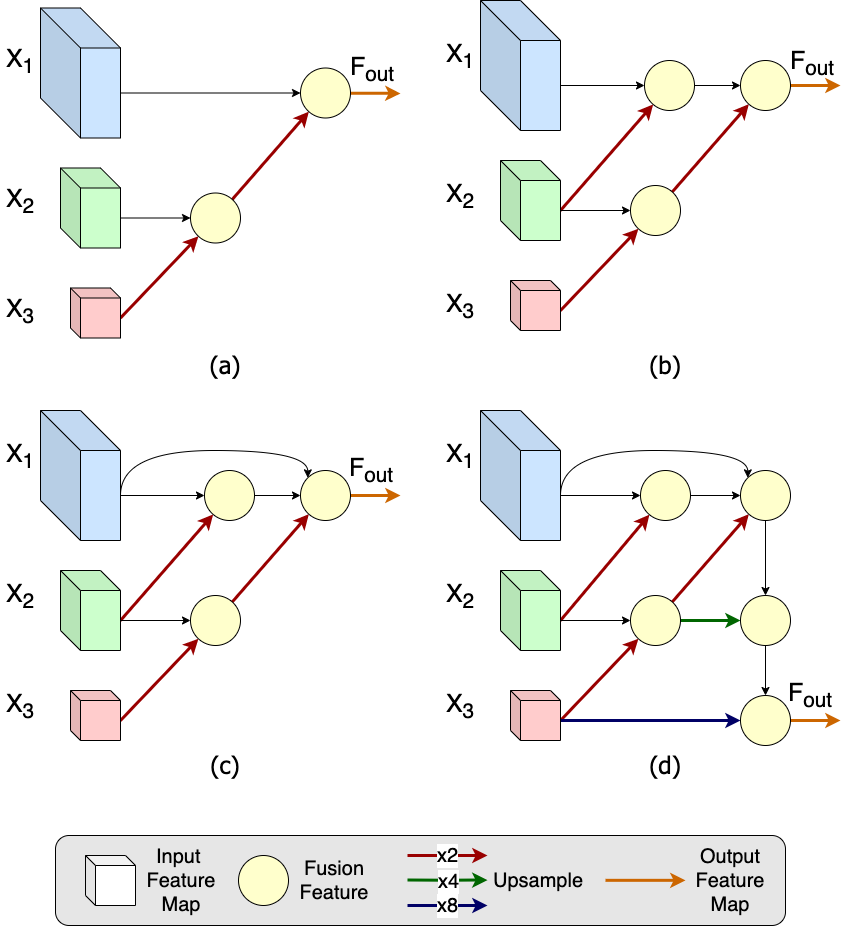}
        \end{center}
        \caption{Different feature aggregation schemes as shown in \cite{zhang2020multi}. \textbf{(a)} Long skip connection (LSC); \textbf{(b)} Iterative Deep Aggregation (IDA); \textbf{(c)} Dense Iterative Aggregation (DIA); \textbf{(d)} Dense Hierarchical Aggregation (DHA).}
        \label{fig:fa}
    \end{figure}
\end{center}

A high polyp miss rate is often associated with small and flat polyps (whose perimeters are below $10 mm$)~\cite{kim2017miss}. In order to detect these polyps, it is important to obtain high-resolution features from multiple image scales. Feature fusion (or aggregation) is a well-studied technique to achieve this for CNNs, in which feature maps from different scales are fused to form a multi-scale feature map. Figure \ref{fig:fa} demonstrates four feature aggregation techniques, in order of increasing complexity \cite{zhang2020multi}.

Long Skip Connection (LSC) illustrated in Figure~\ref{fig:fa}a is an early aggregation scheme used by segmentation networks such as UNet~\cite{ronneberger2015u}, UNet++~\cite{zhou2019unet++}, and Attention UNet~\cite{oktay2018attention}. Given feature maps $X_1$, $X_2$, $X_3$, higher-level feature maps are upsampled and fused with their adjacent low-level features by a long skip connection, gradually restoring the spatial information. Each fusion module includes a concatenation layer and a convolutional layer with kernel size $3\times3$.

LSC is a well-tested and straightforward technique, but it is not without limitations. Zhang et al. \cite{zhang2020multi} proposed three alterations to produce higher-quality features, as described below.

Iterative Deep Aggregation (IDA) depicted in Figure~\ref{fig:fa}b produces finer feature maps by using multiple iterative convolutions for a single scale.

Inspired by DenseNet, Dense Iterative Aggregation (DIA) introduces dense skip connections to the iterative convolutions in IDA (see Figure~\ref{fig:fa}c). This addition ensures maximum information flow and reduces overfitting.

Dense Hierarchical Aggregation (DHA) further enhances semantic information by re-combining the output with high-level feature maps as shown in Figure~\ref{fig:fa}d.

From these observations, in our \ModelName, we propose to apply feature aggregation to the outputs of RFB modules at different scales. We examine variants with each of the aforementioned aggregation schemes, namely \LSC, \IDA, \DIA, and \DHA, in section \ref{sec:experiment}.

\subsection{Loss Function and Auxiliary Training}
We aim to exploit the information from data with the undefined labels for training in a way similar to \cite{lan2021neounet}. The intuition is that while these data do not provide
information for deciding on neoplasm class, they still can provide some semantic meaning from the data for the segmentation.
% any neoplasm information, they still have segmentation data.

To training \ModelName, we propose the loss function $\mathcal{L}_{total}$ consisting of two components: a main loss $\mathcal{L}_{main}$ associated with the main task of neoplasm segmentation, and an auxiliary loss $\mathcal{L}_{aux}$ associated with the auxiliary task of polyp segmentation. The total loss can be expressed as follows:

\begin{equation}
    \label{eq:total_loss}
    \mathcal{L}_{total} =  \mathcal{L}_{main} + \mathcal{L}_{aux}
\end{equation}

The main loss $\mathcal{L}_{main}$ drives the model toward making accurate class-specific segmentation. The pixels with undefined labels are excluded when calculating $\mathcal{L}_{main}$. The auxiliary loss $\mathcal{L}_{aux}$ drives the model toward making accurate foreground-background segmentation, in which all pixels are used for training.

To investigate this training strategy, we iterate through three different ways to incorporate it into \ModelName, which creates three different variants: Single-headed Binary \ModelName (SB-\ModelName), Single-headed Trinary \ModelName (ST-\ModelName), and Multi-headed \ModelName (the final version of \ModelName).

\subsubsection{Single-headed Binary \ModelName}
Our first variant, SB-\ModelName (Figure~\ref{fig:model}a), uses the same loss formulation as the NeoUNet model proposed in \cite{lan2021neounet}. The final output layer produces two binary segmentation maps, one for the neoplastic class and another for the non-neoplastic class. Losses are calculated separately for each map and then averaged. The main loss is a combination of Binary Cross Entropy and Focal Tversky loss \cite{abraham2019novel} as follows:
\begin{align}
     & \mathcal{L}_{main} = \mathcal{L}_{main}^{neo} + \mathcal{L}_{main}^{non} \nonumber                               \\
     & = \mathcal{L}_{BCE}(P_{main}^{neo}, G_{main}^{neo}) + \mathcal{L}_{FT}(P_{main}^{neo}, G_{main}^{neo}) \nonumber \\
     & +  \mathcal{L}_{BCE}(P_{main}^{non}, G_{main}^{non}) + \mathcal{L}_{FT}(P_{main}^{non}, G_{main}^{non})
    \label{eq:main_loss_sb}
\end{align}
where $P_{main}^{neo}$ and $P_{main}^{non}$ denote the prediction maps for the neoplastic and non-neoplastic classes, respectively; $G_{main}^{neo}$ and $G_{main}^{non}$ are ground truths. We choose the Focal Tversky loss for $\mathcal{L}_{main}$ to alleviate class imbalance due to small amount of non-neoplastic polyp pixels as shown later in Figure~\ref{fig:data_dist}.

The auxiliary loss is a combination of Binary Cross Entropy and Tversky loss, which uses an auxiliary polyp segmentation map inferred from the two binary class-specific maps:
\begin{equation}
    \label{eq:aux_loss_sb}
    \mathcal{L}_{aux} =  \mathcal{L}_{BCE}(P_{aux}^{polyp}, G_{aux}^{polyp}) + \mathcal{L}_{T}(P_{aux}^{polyp}, G_{aux}^{polyp})
\end{equation}
where $P_{aux}^{polyp}$ and $G_{aux}^{polyp}$ denote the auxiliary polyp prediction map and the corresponding ground truth.

The auxiliary polyp prediction map $P_{aux}^{polyp}$ is inferred using element-wise max:
\begin{equation}
    P_{aux}^{polyp} = \max(P_{main}^{neo}, P_{main}^{non})
    \label{eq:mapconvert}
\end{equation}

%Figure~\ref{fig:model}a illustrates how SB-\ModelName generates the output maps.

\subsubsection{Single-headed Trinary \ModelName}
While the method using binary map \cite{lan2021neounet} yielded promising classification results, we found that a lighter model can benefit from imposing an additional constraint on the outputs.
Specifically, when the model outputs two separate class-specific maps for one image, it may make both maps have high prediction values for the same pixel. Therefore, a class constraint is needed to ensure that one class is chosen for each pixel. In our second variant, ST-\ModelName (Figure~\ref{fig:model}b), we use a 3-channel output map $P_{main}^{trinary}$, denoting the probabilities for the neoplastic, non-neoplastic, and background class, respectively. A softmax activation is used on the channel dimension, meaning each pixel may only belong to one class.

The main loss is a combination of Categorical Cross Entropy and Focal Tversky loss as follows:

\begin{align}
    \label{eq:main_loss}
    \mathcal{L}_{main} & = \mathcal{L}_{CCE}(P_{main}^{trinary}, G_{main}^{trinary}) \nonumber \\
                       & + \mathcal{L}_{FT}(P_{main}^{trinary}, G_{main}^{trinary})
\end{align}
where $G_{main}^{trinary}$ is the trinary ground truth.

Similarly, the auxiliary loss is a combination of Categorical Cross Entropy and Tversky loss:

\begin{equation}
    \label{eq:aux_loss}
    \mathcal{L}_{aux} =  \mathcal{L}_{CCE}(P_{aux}^{polyp}, G_{aux}^{polyp}) + \mathcal{L}_{T}(P_{aux}^{polyp}, G_{aux}^{polyp})
\end{equation}
Here we apply element-wise max to the two channels of the map $P_{main}^{trinary}$, which correspond to the neoplastic and non-neoplastic classes, to produce the auxiliary polyp segmentation map $P_{aux}^{polyp}$.

%Figure~\ref{fig:model}b illustrates how ST-\ModelName generates the output map.

\subsubsection{Multi-headed \ModelName}
For our final version of \ModelName, we further evolve the use of auxiliary loss by adding an auxiliary segmentation branch.
Auxiliary training is the process of jointly learning a \textit{side} or \textit{auxiliary} task to enhance the main task's performance \cite{chennupati2019auxnet,zhang2020auxiliary}. This idea is similar to multi-task learning, except the auxiliary branch does not activate in the inference phase.
%TODO: có công thức hàm loss cho Multi-headed không? trong khi Single-head thì có Eq. (2) và (3)

The auxiliary branch uses an identical (with separate weights) feature aggregation module but outputs a binary segmentation map instead of a 3-channel multi-class map. The main loss and auxiliary loss are calculated exactly as in Eq.~(\ref{eq:main_loss}) and Eq.~(\ref{eq:aux_loss}), respectively. However, there is no conversion between multi-class and polyp segmentation maps as in Eq.~(\ref{eq:mapconvert}). Instead, the main loss is calculated with the 3-channel multi-class map, and the auxiliary loss is calculated with the output map from the auxiliary branch. Intuitively, we believe the model will benefit from adding an auxiliary network branch since it strengthens the supervised signal and betters the optimization process during training the model. Figure~\ref{fig:model}c describes how \ModelName incorporates the auxiliary branch for training in our BlazeNeo.

% \begin{center}
%     \begin{figure}[th]
%         \begin{center}
%             \includegraphics[scale=.35]{images/auxiliary_learning.png}
%         \end{center}
%         \caption{An auxiliary training strategy for \ModelName. The auxiliary task serves as a regularization factor for faster convergence and boosts the performance for the main task.}
%         \label{fig:auxiliary_learning}
%     \end{figure}
% \end{center}

\section{Experiments}
\label{sec:experiment}

\subsection{Benchmark dataset}

\begin{figure*}[ht!]
    \begin{center}
        \includegraphics[width=\textwidth]{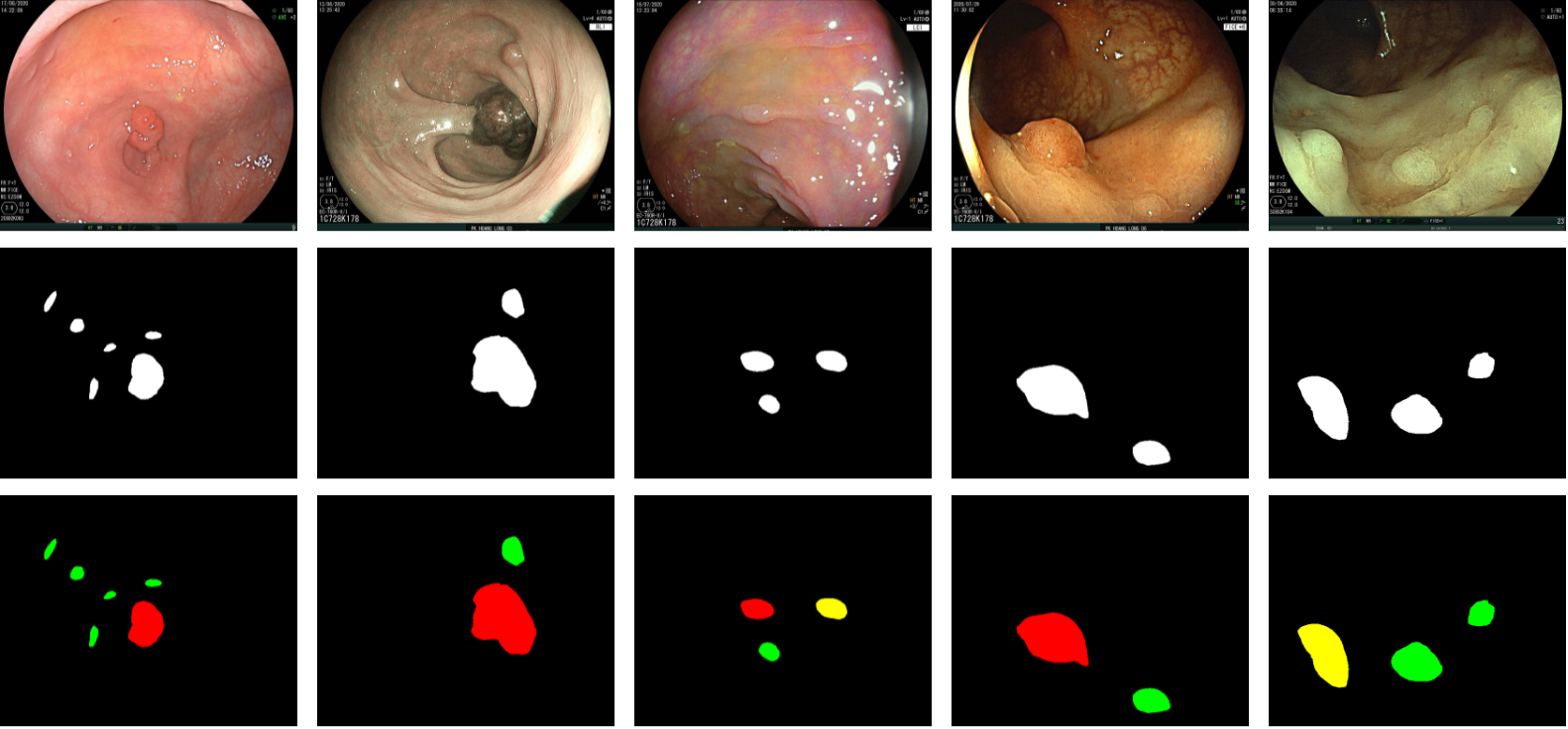}
    \end{center}
    \caption{Some examples from the NeoPolyp dataset. The first row displays original images from the dataset. The second row shows the ground truths for polyp segmentation. The last row shows the ground truths for neoplasm segmentation, where some polyps are undefined and marked by yellow color. From left to right, the color modes are WLI, BLI, LCI, FICE, and FICE, respectively.}
    \label{fig:polyp:examples}
\end{figure*}

\begin{center}
    \begin{figure*}[ht!]
        \subcaptionbox{Polyp-wise distribution}{
            \includegraphics[width=0.48\textwidth]{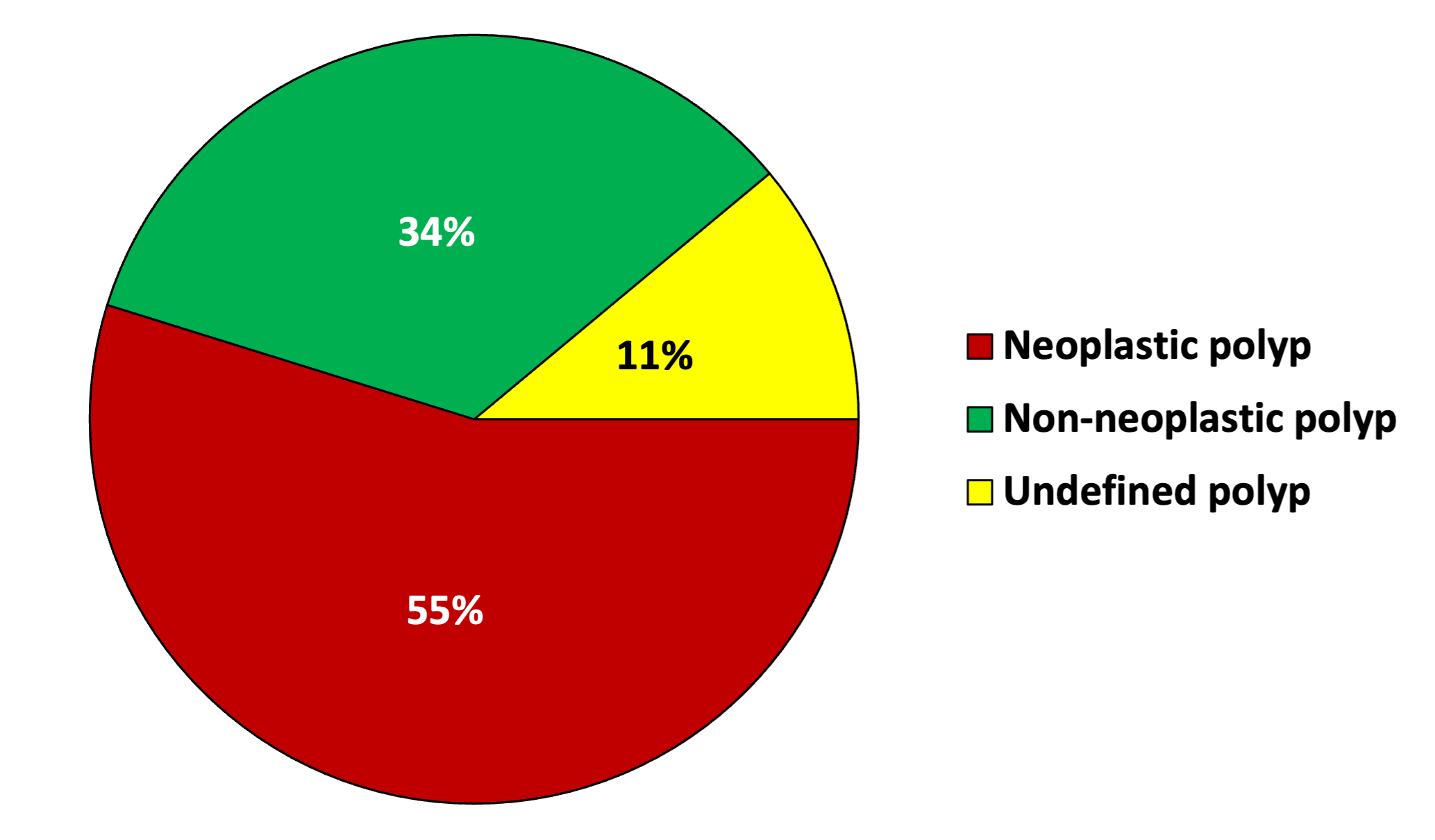}
        }
        \subcaptionbox{Pixel-wise distribution}{
            \includegraphics[width=0.48\textwidth]{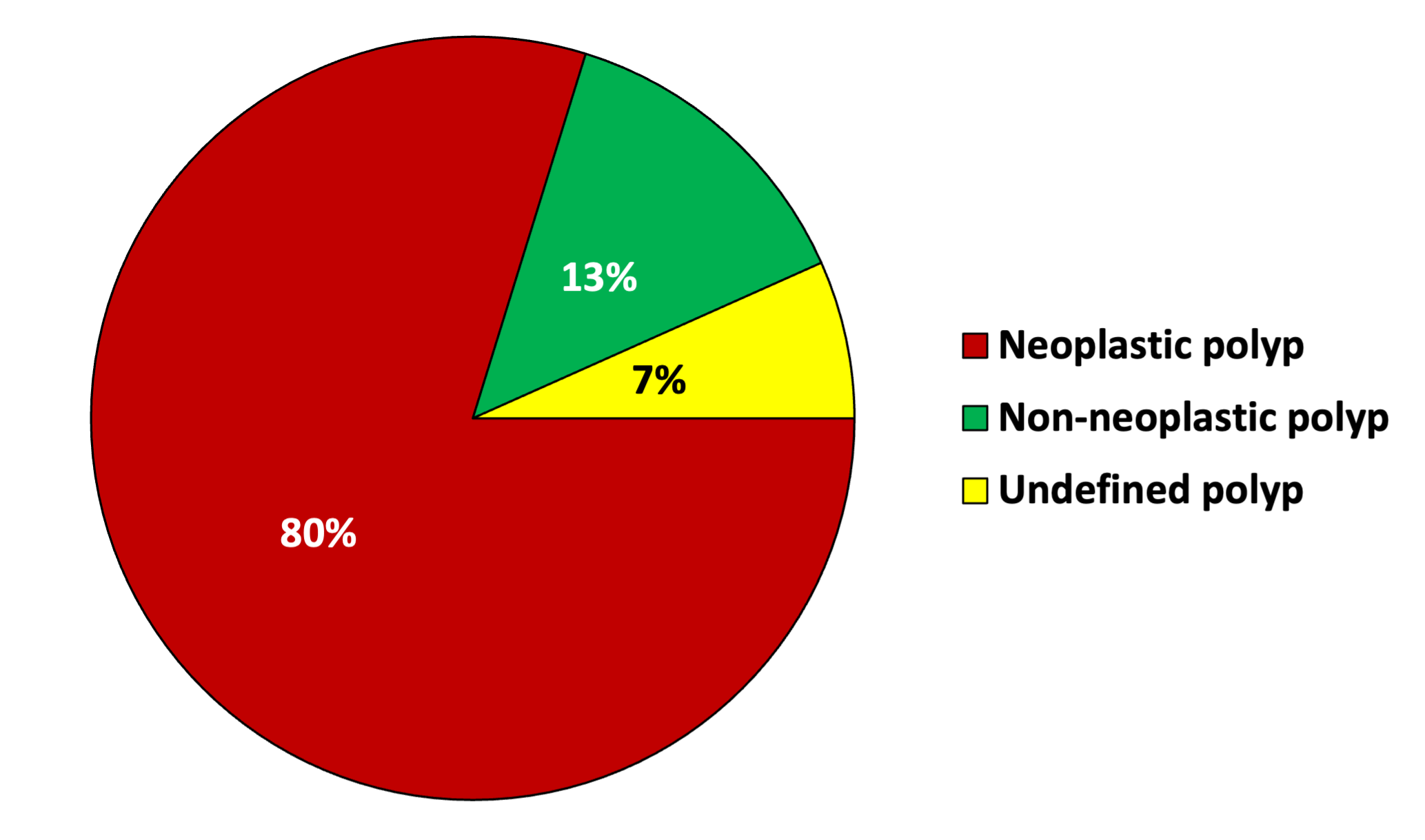}
        }
        \caption{Data distribution of polyp class labels in the \DatasetName dataset. In the pixel-wise distribution on the right, percentages are calculated on polyp pixels only (not including background pixels.)}
        \label{fig:data_dist}
    \end{figure*}
\end{center}

We use the \DatasetName dataset as introduced in \cite{lan2021neounet} to train and benchmark the proposed \ModelName{}. The dataset consists of 7,466 annotated endoscopic images captured directly during endoscopic recording and includes all four lighting modes: WLI (White Light Imaging), FICE (Flexible spectral Imaging Color Enhancement), BLI (Blue Light Imaging), and LCI (Linked Color Imaging). \DatasetName is split into a training set of 5,966 images and a test set of 1,500 images. Some examples of the \DatasetName dataset are shown in Figure~\ref{fig:polyp:examples}. For comparison with baseline models, we also use the \CleanDatasetName dataset, which does not contain any polyps with undefined class labels. This dataset consists of 5,277 training images and 1,353 test images.

In practice, most non-neoplastic polyps are small, and the endoscopists can immediately remove them without the need to capture images or take a biopsy for lesions less than 5mm for post-checking. Due to that reason, neoplastic polyps take up a majority of the polyps present in \DatasetName (see Figure~\ref{fig:data_dist}). The number of neoplastic, non-neoplastic, and undefined polyps are 5113, 3185, and 1031, respectively. However, if we look at the pixel-wise level as shown in Figure~\ref{fig:data_dist}b, we can observe a strong data imbalance between the three classes. The number of neoplastic polyp pixels takes up to 80\% among all polyp pixels in the dataset. Meanwhile, these numbers for non-neoplastic and undefined classes are 13\% and 7\%, respectively. This data imbalance, combined with the inherent challenges of PSND, creates a difficult benchmark for models to overcome.

\subsection{Experiment setup}
\begin{center}
    \begin{figure*}[ht]
        % \begin{center}
        %     \includegraphics[scale=.7]{images/aux-effectiveness-non.png}
        % \end{center}
        \subcaptionbox{Non-neoplastic class}{
            \includegraphics[width=0.5\textwidth]{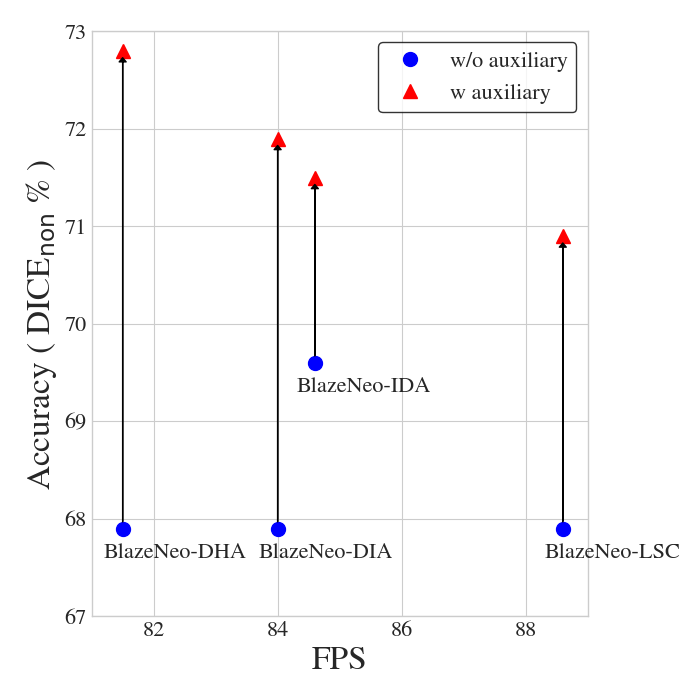}
        }
        \subcaptionbox{Neoplastic class}{
            \includegraphics[width=0.5\textwidth]{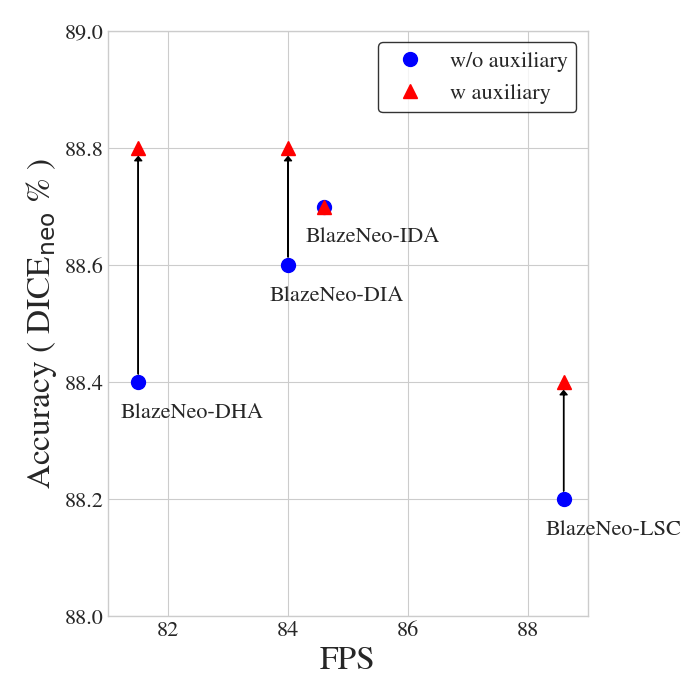}
        }
        \caption{Dice scores and FPS for different \ModelName variations, with and without auxiliary training. Red triangles denote results with auxiliary training, while blue circles are those without auxiliary training.}
        \label{fig:aux-effectiveness}
    \end{figure*}
\end{center}

% \begin{figure*}
%     \begin{subfigure}[th!]{0.25\textwidth}
%             \centering
%             \includegraphics[width=.99\linewidth]{images/lsc_loss.jpg}
%             \caption{\LSC}
%             \label{fig:gull}
%     \end{subfigure}%
%     \begin{subfigure}[th!]{0.25\textwidth}
%             \centering
%             \includegraphics[width=.99\linewidth]{images/ida_loss.jpg}
%             \caption{\IDA}
%             \label{fig:gull2}
%     \end{subfigure}%
%     \begin{subfigure}[th!]{0.25\textwidth}
%             \centering
%             \includegraphics[width=.99\linewidth]{images/dia_loss.jpg}
%             \caption{\DIA}
%             \label{fig:tiger}
%     \end{subfigure}%
%     \begin{subfigure}[th!]{0.25\textwidth}
%             \centering
%             \includegraphics[width=.99\linewidth]{images/dha_loss.jpg}
%             \caption{\DHA}
%             \label{fig:mouse}
%     \end{subfigure}
%     \caption{The training loss graph of different \ModelName models with and without auxiliary training strategy.}\label{fig:aux-loss}
% \end{figure*}

Our experiments include several ablation studies to verify the effectiveness of each component in \ModelName, a comparison against several polyp segmentation methods, and benchmarks on the NVIDIA Jetson AGX Xavier developer kit \footnote{https://developer.nvidia.com/embedded/jetson-agx-xavier-developer-kit}, which closely resembles real deployments. The Jetson device is configured to run at MAXN power mode.

We oversample non-neoplastic polyps to account for class imbalance in the \DatasetName dataset, as addressed by \cite{lan2021neounet}. Images containing non-neoplastic polyps are duplicated such that $P_{non} \approx P_{neo}$, where $P_{non}$ and $P_{neo}$ are the number of pixels containing in  non-neoplastic and neoplastic polyps, respectively.

The models are trained using Stochastic Gradient Descent (SGD) with Nesterov momentum and an initial learning rate of $0.001$. The learning rate is adjusted according to a combination of linear warmup and a cosine annealing schedule.

Images used for training are at 3 different scales: $256 \times 256$, $352 \times 352$ and $512 \times 512$.

During training, augmentations including random scaling, rotation, horizontal/vertical flip, motion blur, and color jittering are added to improve generality. These augmentations are performed on-the-fly with a probability of $0.5$.

\ModelName and other baseline models are implemented in Python 3.7 using the PyTorch framework.

\subsection{Evaluation metrics}
Commonly used metrics Dice score and IoU score are employed to measure the model's output quality. They are evaluated on three classes: neoplastic polyp, non-neoplastic polyp, and generic polyp (same as polyp segmentation). Dice and IoU are calculated pixel-wise on the entire test set (micro-averaged). Equations (\ref{eq:dice_score}) and (\ref{eq:iou_score}) describe how these metrics are calculated.

\begin{equation}
    \label{eq:dice_score}
    Dice_{c} = \frac{ 2\times \sum_{i\in{I}}^{} u_{i}^{c} v_{i}^{c} }{ \sum_{i\in{I}}^{} u_{i}^{c} + \sum_{i\in{I}}^{} v_{i}^{c}}
\end{equation}

\begin{equation}
    \label{eq:iou_score}
    IoU_{c} = \frac{ \sum_{i\in{I}}^{} u_{i}^{c} v_{i}^{c} }{ \sum_{i\in{I}}^{} u_{i}^{c} + \sum_{i\in{I}}^{} v_{i}^{c} - \sum_{i\in{I}}^{} u_{i}^{c} v_{i}^{c} }
\end{equation}
where $i \in I$ denotes a prediction pixel within the entire test set. $u_{i}^{c} = 1$ if the model predicts pixel $i$ to have class $c$, and $0$ otherwise. Similarly, $v_{i}^{c} = 1$ if the ground truth map states that pixel $i$ has class $c$, and $0$ otherwise. For neoplastic and non-neoplastic class evaluation, $I$ does not include undefined neoplasm pixels.

We also evaluate each model's inference speed using the number of processed frames per second (FPS). This metric is measured by running each model with a batch size of 1 on 100 colonoscopy images. FPS is measured on a Google Colaboratory instance with an NVIDIA Tesla V100 GPU when not specified otherwise.

Finally, we log each model's number of parameters and floating-point operations (measured in GFLOPs) to evaluate their size and complexity.

\section{Results and Discussion}
\subsection{Ablation Study}
\subsubsection{The effectiveness of different output architectures}
\begin{table*}[ht!]
    \centering
    \caption{Performance metrics on the \DatasetName test set for the three variants of BlazeNeo using the same DHA feature aggregation scheme}
    \label{tab:exp3_2}
    \begin{tabular}{@{} c| c c c c c c @{}}
        \toprule
        
        \multicolumn{1}{c|}{Model} & \multicolumn{1}{c}{$\text{Dice}_{seg}$} & \multicolumn{1}{c}{$\text{IoU}_{seg}$} & \multicolumn{1}{c}{$\text{Dice}_{non}$} & \multicolumn{1}{c}{$\text{IoU}_{non}$} & \multicolumn{1}{c}{$\text{Dice}_{neo}$} & \multicolumn{1}{c}{$\text{IoU}_{neo}$} \\ \midrule \midrule
        SB-\ModelName              & 0.866                                   & 0.764                                  & 0.683                                   & 0.518                                  & 0.862                                   & 0.758                                  \\
        ST-\ModelName              & 0.874                                   & 0.777                                  & 0.671                                   & 0.505                                  & 0.865                                   & 0.762                                  \\
        \ModelName                 & \textbf{0.901}                          & \textbf{0.820}                         & \textbf{0.728}                          & \textbf{0.572}                         & \textbf{0.888}                          & \textbf{0.800}                         \\
        \bottomrule
    \end{tabular}
\end{table*}

Firstly, we compare our three \ModelName variants shown in Figure \ref{fig:model} to evaluate the effectiveness of different output architectures. Table \ref{tab:exp3_2} shows performance metrics for each variant on the \DatasetName test set. Here we use the same DHA feature aggregation scheme for all three of \ModelName. The final \ModelName model with multi heads achieves the best results on all metrics. Notably, this final variant outperforms the other two variants by over $6\%$ in IoU score for the non-neoplastic class. This shows the effectiveness of the auxiliary branch, which helps anchor segmentation performance and makes use of undefined labels. We also see that ST-\ModelName outperforms SB-\ModelName, which justifies our use of trinary output in the final variant.

\subsubsection{The effectiveness of different feature aggregation schemes}
This experiment investigates the use of four different feature aggregation schemes in \ModelName{}: Long Skip Connection (LSC), Iterative Deep Aggregation (IDA), Dense Iterative Aggregation (DIA), and Dense Hierarchical Aggregation (DHA).

Table~\ref{tab:exp1} shows performance metrics for each model variation on the \DatasetName dataset. We can see that \DHA produces the highest values of Dice and IoU on all classes. This is what we expected, as DHA is the most complex aggregation mechanism that preserves a lot of high-resolution features. However, this improvement comes at a cost as \DHA is also the slowest variation at only 81.5 FPS, compared to the fastest \LSC at 88.6 FPS. Interestingly, the nested IDA aggregation scheme performs worse than basic LSC in the segmentation task. This is alleviated in DIA and DHA with the use of skip connections.

\begin{table*}[ht!]
    \centering
    \caption{Performance metrics on the \DatasetName test set for \ModelName with different feature aggregation schemes}
    \label{tab:exp1}
    \begin{tabular}{@{} c | c c c c c c | c@{}}
        \toprule
        
        \multicolumn{1}{c | }{Method} & \multicolumn{1}{c}{$\text{Dice}_{seg}$} & \multicolumn{1}{c}{$\text{IoU}_{seg}$} & \multicolumn{1}{c}{$\text{Dice}_{non}$} & \multicolumn{1}{c}{$\text{IoU}_{non}$} & \multicolumn{1}{c}{$\text{Dice}_{neo}$} & \multicolumn{1}{c|}{$\text{IoU}_{neo}$} & \multicolumn{1}{c}{FPS} \\ \midrule \midrule
        \LSC                          & 0.897                                   & 0.814                                  & 0.709                                   & 0.550                                  & 0.884                                   & 0.792                                   & \textbf{88.6}           \\
        \IDA                          & 0.890                                   & 0.803                                  & 0.715                                   & 0.557                                  & 0.887                                   & 0.798                                   & 84.6                    \\
        \DIA                          & 0.897                                   & 0.814                                  & 0.719                                   & 0.562                                  & \textbf{0.888}                          & \textbf{0.800}                          & 84.0                    \\
        \DHA                          & \textbf{0.901}                          & \textbf{0.820}                         & \textbf{0.728}                          & \textbf{0.572}                         & \textbf{0.888}                          & \textbf{0.800}                          & 81.5                    \\
        \bottomrule
    \end{tabular}
\end{table*}

\subsubsection{The effectiveness of auxiliary training}

This experiment looks into the effectiveness of enabling and disabling auxiliary training for each variation of \ModelName, namely \LSC, \IDA, \DIA, and \DHA.

Figure~\ref{fig:aux-effectiveness} shows that auxiliary training generally improves output quality. In fact, without auxiliary segmentation learning, \DIA and \DHA achieve even lower accuracy than \IDA. The non-neoplastic class benefits the most from auxiliary training, improving \LSC, \IDA, \DIA, and \DHA by 3\%, 1.9\%, 4\%, and 4.9\%, respectively.

% Follow~\cite{liebel2018auxiliary}, we evaluate the performance at intermediate optimization states.
% Figure~\ref{fig:aux-loss} shows the training loss curves of the main task.
% As expected, there is a significant difference in the learning capacity of the models when we activate and deactivate auxiliary learning. The auxiliary task serves as regularization and helps the main task converge faster.
% It justifies that the auxiliary training strategy can provide better convergence for the networks in general. More importantly, it boosts model performance without introducing extra computation during the inference stage.

\begin{table*}[ht!]
    \centering
    \caption{Performance metrics for \DHA when training on \DatasetName and \CleanDatasetName, measured on the \DatasetName test set}
    \label{tab:exp3}
    \begin{tabular}{@{} c| c c c c c c @{}}
        \toprule
        
        \multicolumn{1}{c|}{Training dataset} & \multicolumn{1}{c}{$\text{Dice}_{seg}$} & \multicolumn{1}{c}{$\text{IoU}_{seg}$} & \multicolumn{1}{c}{$\text{Dice}_{non}$} & \multicolumn{1}{c}{$\text{IoU}_{non}$} & \multicolumn{1}{c}{$\text{Dice}_{neo}$} & \multicolumn{1}{c}{$\text{IoU}_{neo}$} \\ \midrule \midrule
        \CleanDatasetName                     & 0.900                                   & 0.818                                  & 0.714                                   & 0.555                                  & 0.884                                   & 0.792                                  \\
        \DatasetName                          & \textbf{0.901}                          & \textbf{0.820}                         & \textbf{0.728}                          & \textbf{0.572}                         & \textbf{0.888}                          & \textbf{0.799}                         \\
        
        \bottomrule
    \end{tabular}
\end{table*}

\subsubsection{The effectiveness of including undefined polyps}
This experiment examines the effectiveness of using undefined neoplasm pixels via the auxiliary module. Table~\ref{tab:exp3} shows performance metrics for \DHA, the best-performing \ModelName model, when trained on the \DatasetName dataset (which contains undefined polyps) and the \CleanDatasetName dataset (which does not contain undefined polyps). Results show that the addition of these undefined pixels leads to improvements across the board, especially for the non-neoplastic class.

\subsection{Comparison with state-of-the-art models}
\subsubsection{Quantitative comparison}
We compare the performance of \DHA, the best-performing \ModelName model, with seven state-of-the-art models for the polyp segmentation and PSND problem: U-Net \cite{ronneberger2015u}, ColonSegNet \cite{jha2021real}, DDANet \cite{tomar2020ddanet}, DoubleUNet \cite{jha2020doubleu},  HarDNet-MSEG \cite{huang2021hardnet}, PraNet \cite{fan2020pranet}, and NeoUNet \cite{lan2021neounet}. Except for NeoUNet, the models mentioned above do not handle undefined neoplasm pixels during training. Thus, in the interest of a fair comparison, we use the \CleanDatasetName dataset for this experiment, which does not contain undefined polyps. Results are shown in Table \ref{tab:compare_sota}.

\begin{table*}[ht!]
    \centering
    \caption{Performance metrics of different models on the \CleanDatasetName test set}
    \label{tab:compare_sota}
    \begin{tabular}{@{} c | c c c c c c | c c c@{}}
        \toprule
        
        \multicolumn{1}{c | }{Method}        & \multicolumn{1}{c}{$\text{Dice}_{seg}$} & \multicolumn{1}{c}{$\text{IoU}_{seg}$} & \multicolumn{1}{c}{$\text{Dice}_{non}$} & \multicolumn{1}{c}{$\text{IoU}_{non}$} & \multicolumn{1}{c}{$\text{Dice}_{neo}$} & \multicolumn{1}{c | }{$\text{IoU}_{neo}$} & \multicolumn{1}{c}{FPS} & \multicolumn{1}{c}{Parameters} & \multicolumn{1}{c}{GFLOPs} \\ \midrule
        \midrule
        ColonSegNet \cite{jha2021real}       & 0.738                                   & 0.585                                  & 0.505                                   & 0.338                                  & 0.732                                   & 0.577                                     & 44.9                    & 5,010,000                      & 64.84                      \\
        U-Net \cite{ronneberger2015u}        & 0.785                                   & 0.646                                  & 0.525                                   & 0.356                                  & 0.773                                   & 0.631                                     & 69.6                    & 31,043,651                     & 103.59                     \\
        DDANet \cite{tomar2020ddanet}        & 0.813                                   & 0.684                                  & 0.578                                   & 0.406                                  & 0.802                                   & 0.670                                     & 46.2                    & 6,840,000                      & 31.45                      \\
        DoubleU-Net \cite{jha2020doubleu}    & 0.837                                   & 0.720                                  & 0.621                                   & 0.450                                  & 0.832                                   & 0.712                                     & 43.2                    & 18,836,804                     & 83.62                      \\
        HarDNet-MSEG \cite{huang2021hardnet} & 0.883                                   & 0.791                                  & 0.659                                   & 0.492                                  & 0.869                                   & 0.769                                     & \underline{77.1}        & \underline{17,424,031}         & \underline{11.38}          \\
        PraNet \cite{fan2020pranet}          & 0.895                                   & 0.811                                  & 0.705                                   & 0.544                                  & 0.873                                   & 0.775                                     & 55.6                    & 30,501,341                     & 13.11                      \\
        NeoUNet \cite{lan2021neounet}        & \textbf{0.911}                          & \textbf{0.837}                         & \textbf{0.720}                          & \textbf{0.563}                         & \textbf{0.889}                          & \textbf{0.800}                            & 68.3                    & 38,288,397                     & 39.88                      \\
        \DHA (Ours)                          & \underline{0.904}                       & \underline{0.825}                      & \underline{0.717}                       & \underline{0.559}                      & \underline{0.885}                       & \underline{0.792}                         & \textbf{81.5}           & \textbf{17,143,324}            & \textbf{11.06}             \\
        
        \bottomrule
    \end{tabular}
\end{table*}

We can see that while NeoUNet remains the most accurate model, \ModelName is a close second, with a difference of less than $1\%$ on most accuracy metrics. At the same time, \ModelName is a much more lightweight and faster model. Compared to NeoUNet, \ModelName achieves higher FPS ($81.5$ versus $68.3$), has half as many parameters ($17,143,324$ versus $38,288,397$), and lower GFLOPs ($11.06$ versus $39.88$). Notably, \DHA is faster and more accurate than HarDNet-MSEG, while NeoUNet is slower than HarDNet-MSEG and U-Net. We attribute the speed improvement of \ModelName over HarDNet-MSEG to the different decoder designs, especially the use of the small RFB module.

Although achieving a high overall dice score, our \ModelName is not without limitations. The dice and IoU scores of \ModelName are still low for non-neoplastic polyps, and both of them are significantly reduced when the model is converted into the INT8 precision. The reason may be due to the small size of non-neoplastic polyps. When the weights are converted into integers ranging from -128 to 127 in the INT8 precision, the model seems to lose the capacity of representing detailed information and, therefore, is easier to miss small objects in the input images. 

\subsubsection{Qualitative comparison}
\begin{center}
    \begin{figure*}[th!]
        \begin{center}
            \includegraphics[scale=.21]{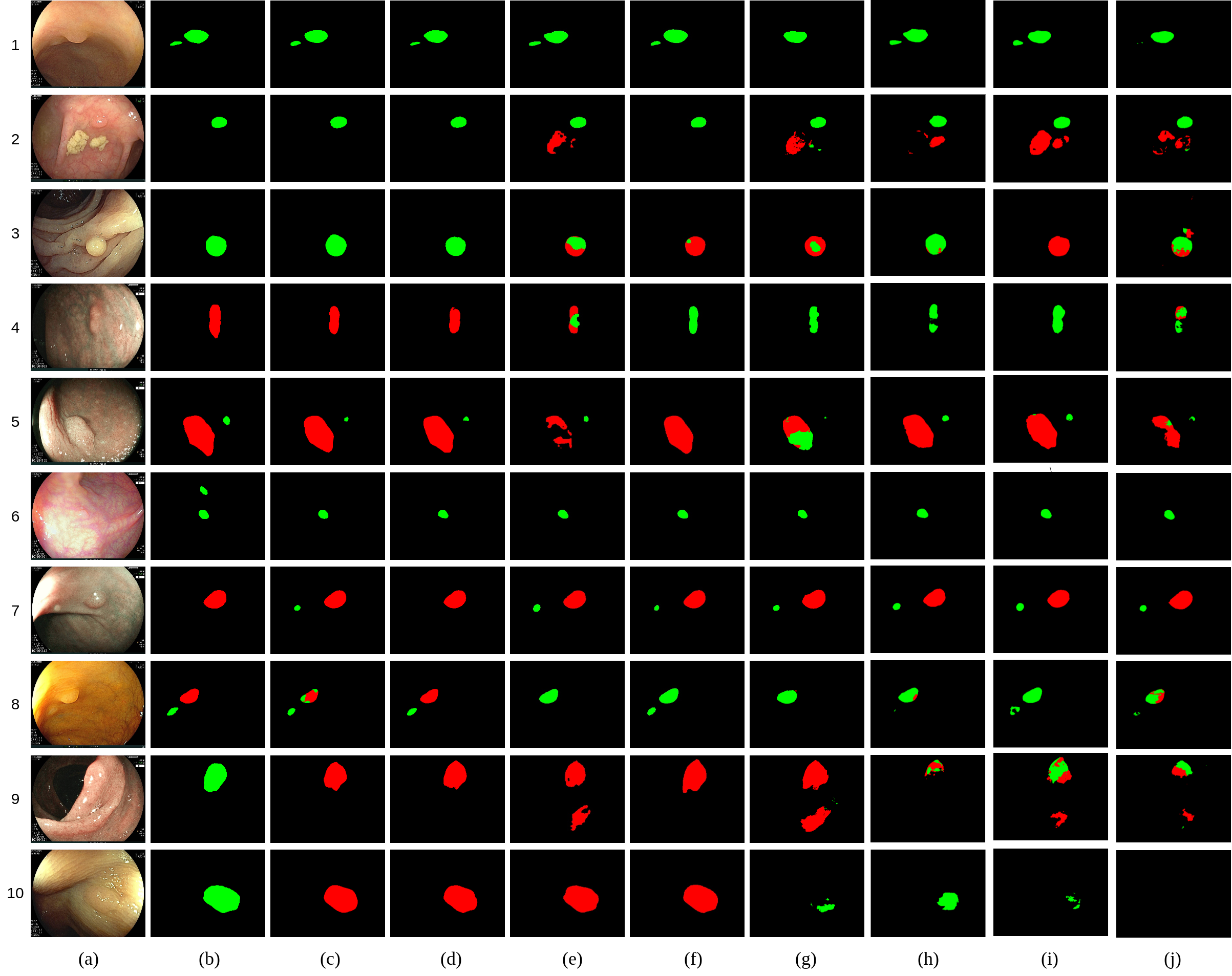}
        \end{center}
        \caption{Qualitative comparison of the proposed method with other baseline methods: (a) image, (b) ground truth, (c) BlazeNeo (Ours), (d) NeoUNet, (e) PraNet, (f) HarDNet-MSEG, (g) UNet, (h) DoubleU-Net, (i) DDANet, and (j) ColonSegNet.}
        \label{fig:good}
    \end{figure*}
\end{center}

Figure~\ref{fig:good} shows examples outputs of \DHA and other baseline models. Overall, \DHA produces the most accurate segmentation and classification results for different types of polyps.

The first five rows of Figure~\ref{fig:good} contain ``easier'' polyp examples. \DHA and NeoUNet perform quite well in these examples, with similarly high accuracy. Meanwhile, PraNet, HarDNet-MSEG, DoubleU-Net, DDANet, U-Net and ColonSegNet produce predictions with less uniformity, with some polyps containing both neoplastic and non-neoplastic regions. For larger polyps such as the 5th row, PraNet is unable to fully segment the area.

The final five rows in Figure~\ref{fig:good} represent more challenging examples, in which all models struggle to provide accurate segmentation masks. This is because non-neoplastic polyps are usually small in size and easier to be miss detected. \DHA and NeoUNet produce fewer false positives in these situations, while U-Net and ColonSegNet create the most inaccurate masks.

The last two rows show two non-neoplastic polyps in BLI and FICE modes. Contributing factors for these struggles in enhanced color modes include their smaller proportions in the training dataset. Furthermore, the endoscopist put the camera scope close to the mucosa in these cases, making the polyps' surface look very clear and sharp such that one can even observe small dots that are glandular holes on the surface. Therefore, these polyps are easily confused with angiogenesis in neoplastic lesions, which is why all models failed in classifying them.

%Contributing factors for these struggles include rarer color modes (e.g., BLI, FICE) in the training set in rows 7 and 10 or abnormal camera angles.

% The last four rows of Figure~\ref{fig:good} contain only neoplastic polyps. HarDNet-MSEG performs poorly on these images, misclassifying neoplastic polyps as non-neoplastic. This can be very costly in actual deployments. In the 7th row of Figure~\ref{fig:good}, only PraNet misses the neoplastic polyp.

\subsection{Benchmark for embedded device}
In this experiment, we apply model compression techniques via the NVIDIA TensorRT 7.1 toolkit \cite{tensorrt} to \ModelName and the baseline models. The compressed models are then benchmarked on the NVIDIA Jetson AGX Xavier developer kit, an embedded computation unit with an NVIDIA GPU specialized for edge AI deployments. This setup is more in-line with deployment scenarios for polyp segmentation and PSND models, i.e., embedded on-site into colonoscopy devices.

Three available precision modes are tested for compressing each model: FP32, FP16, and INT8. Each mode can be seen as a different trade-off level between accuracy and speed. FP32 precision applies techniques such as layer/tensor fusion while keeping parameters as 32-bit floating-point numbers. Hence, this mode provides some speed-up while minimizing accuracy degradation. FP16 precision converts suitable parameters to 16-bit floating-point numbers, greatly reducing model size and latency but is subject to more degradation. Finally, INT8 precision mode quantizes model parameters to 8-bit integers. This precision mode requires an additional calibration procedure to maintain model integrity, which attempts to replicate the original model's output on a small calibration dataset. Despite calibration, INT8 precision is susceptible to a lot more degradation. For this experiment, INT8 calibration is done on a randomized set of images.

Table \ref{tab:embedded} shows performance metrics in different precisions for U-Net, PraNet, HarDNet-MSEG, NeoUNet, and \DHA. We can see a stark difference in FPS when running models on the Jetson AGX Xavier compared to the Google Colab environment. At FP32 precision, the fastest model (\ModelName) achieves only 53.3 FPS on the device, while the slowest model in the Colab environment (PraNet) without any compression still achieves 55.6 FPS. On the other hand, Dice and IoU measures are hardly affected at this precision level. In fact, U-Net and HarDNet-MSEG see improvements up to $1.5\%$ across all metrics after compression. DoubleU-Net and ColonSegNet achieve a slight improvement of about $0.5-1\%$ after compression. DDANet suffers a little from compression, dropping by about $1.5\%$ in segmentation metrics. Finally, PraNet is most affected after compression, dropping by $4.6\%$ in segmentation metrics.
\begin{table*}[ht!]
    \centering
    \caption{Performance metrics of different models with FP16, FP32 and INT8 precision levels on NVIDIA Jetson AGX Xavier}
    \label{tab:embedded}
    \begin{tabular}{@{} c | c | c c c c c c | c @{}}
        \toprule
        
        \multicolumn{1}{c |}{Model} & \multicolumn{1}{c | }{Precision} & \multicolumn{1}{c}{$\text{Dice}_{seg}$} & \multicolumn{1}{c}{$\text{IoU}_{seg}$} & \multicolumn{1}{c}{$\text{Dice}_{non}$} & \multicolumn{1}{c}{$\text{IoU}_{non}$} & \multicolumn{1}{c}{$\text{Dice}_{neo}$} & \multicolumn{1}{c|}{$\text{IoU}_{neo}$} & \multicolumn{1}{c}{FPS} \\ \midrule \midrule
        ColonSegNet@FP32            & FP32                             & 0.745                                   & 0.594                                  & 0.528                                   & 0.359                                  & 0.728                                   & 0.572                                   & 9.4                     \\
        UNet@FP32                   & FP32                             & 0.800                                   & 0.667                                  & 0.537                                   & 0.367                                  & 0.781                                   & 0.641                                   & 10.7                    \\
        DDANet@FP32                 & FP32                             & 0.799                                   & 0.665                                  & 0.557                                   & 0.386                                  & 0.776                                   & 0.635                                   & 16.6                    \\
        DoubleU-Net@FP32            & FP32                             & 0.840                                   & 0.725                                  & 0.627                                   & 0.456                                  & 0.835                                   & 0.717                                   & 10.6                    \\
        HarDNet-MSEG@FP32           & FP32                             & 0.891                                   & 0.804                                  & 0.685                                   & 0.521                                  & 0.871                                   & 0.771                                   & \underline{52.2}        \\
        PraNet@FP32                 & FP32                             & 0.849                                   & 0.738                                  & 0.571                                   & 0.400                                  & 0.844                                   & 0.730                                   & 37.0                    \\
        NeoUNet@FP32                & FP32                             & \textbf{0.909}                          & \textbf{0.832}                         & \textbf{0.725}                          & \textbf{0.568}                         & \textbf{0.893}                          & \textbf{0.806}                          & 25.4                    \\
        \ModelName{}@FP32 (Ours)    & FP32                             & \underline{0.906}                       & \underline{0.828}                      & \underline{0.721}                       & \underline{0.563}                      & \underline{0.887}                       & \underline{0.796}                       & \textbf{53.3}           \\
        \midrule
        ColonSegNet@FP16            & FP16                             & 0.746                                   & 0.594                                  & 0.528                                   & 0.359                                  & 0.728                                   & 0.572                                   & 26.3                    \\
        UNet@FP16                   & FP16                             & 0.800                                   & 0.666                                  & 0.537                                   & 0.367                                  & 0.781                                   & 0.641                                   & 35.5                    \\
        DDANet@FP16                 & FP16                             & 0.799                                   & 0.665                                  & 0.557                                   & 0.386                                  & 0.776                                   & 0.635                                   & 31.8                    \\
        DoubleU-Net@FP16            & FP16                             & 0.840                                   & 0.725                                  & 0.627                                   & 0.456                                  & 0.835                                   & 0.717                                   & 22.9                    \\
        HarDNet-MSEG@FP16           & FP16                             & 0.891                                   & 0.804                                  & 0.685                                   & 0.521                                  & 0.871                                   & 0.771                                   & \underline{118.7}       \\
        PraNet@FP16                 & FP16                             & 0.850                                   & 0.740                                  & 0.572                                   & 0.401                                  & 0.845                                   & 0.731                                   & 95.1                    \\
        NeoUNet@FP16                & FP16                             & \textbf{0.908}                          & \textbf{0.832}                         & \textbf{0.724}                          & \textbf{0.568}                         & \textbf{0.893}                          & \textbf{0.806}                          & 73.4                    \\
        \ModelName{}@FP16 (Ours)    & FP16                             & \underline{0.906}                       & \underline{0.828}                      & \underline{0.721}                       & \underline{0.563}                      & \underline{0.887}                       & \underline{0.796}                       & \textbf{121.9}          \\
        \midrule
        ColonSegNet@INT8            & INT8                             & 0.672                                   & 0.507                                  & 0.456                                   & 0.295                                  & 0.633                                   & 0.463                                   & 41.6                    \\
        UNet@INT8                   & INT8                             & 0.764                                   & 0.618                                  & 0.501                                   & 0.334                                  & 0.753                                   & 0.604                                   & 44.8                    \\
        DDANet@INT8                 & INT8                             & 0.770                                   & 0.626                                  & 0.492                                   & 0.326                                  & 0.748                                   & 0.598                                   & 46.4                    \\
        DoubleU-Net@INT8            & INT8                             & 0.835                                   & 0.717                                  & 0.562                                   & 0.391                                  & 0.830                                   & 0.709                                   & 46.7                    \\
        HarDNet-MSEG@INT8           & INT8                             & 0.810                                   & 0.680                                  & 0.594                                   & 0.422                                  & 0.799                                   & 0.665                                   & \underline{143.7}       \\
        PraNet@INT8                 & INT8                             & 0.817                                   & 0.691                                  & 0.575                                   & 0.404                                  & 0.815                                   & 0.688                                   & 116.7                   \\
        NeoUNet@INT8                & INT8                             & \underline{0.848}                       & \underline{0.736}                      & \underline{0.638}                       & \underline{0.468}                      & \underline{0.848}                       & \underline{0.737}                       & 100.3                   \\
        \ModelName{}@INT8 (Ours)    & INT8                             & \textbf{0.870}                          & \textbf{0.770}                         & \textbf{0.678}                          & \textbf{0.513}                         & \textbf{0.857}                          & \textbf{0.750}                          & \textbf{155.6}          \\
        \bottomrule
    \end{tabular}
\end{table*}

\begin{figure*}[ht!]
    \begin{center}
        \includegraphics[width=0.9\textwidth]{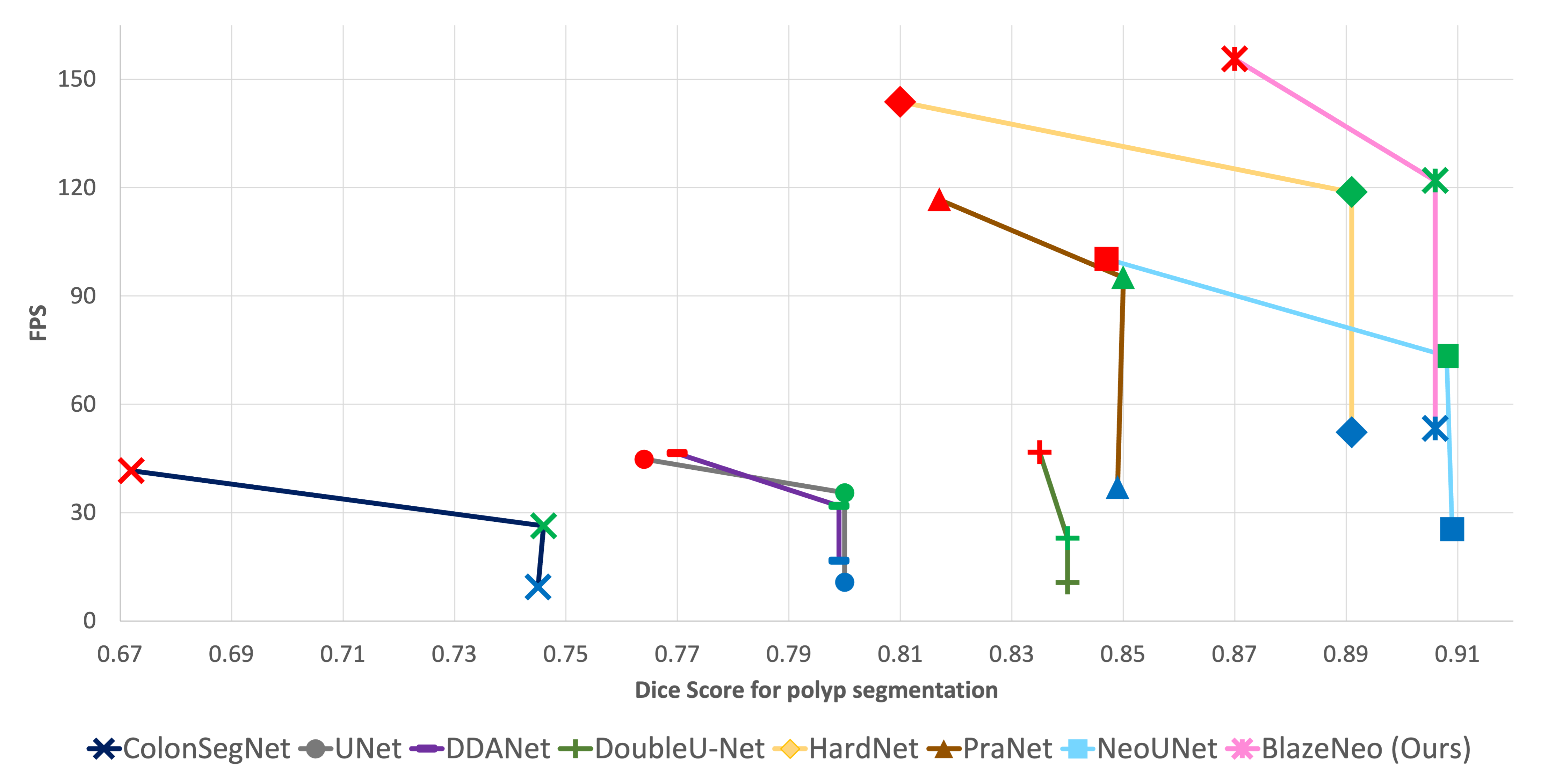}
    \end{center}
    \caption{Comparison between different models in different precisions. Red, green and blue markers denote INT8, FP16 and FP32 precisions, respectively.}
    \label{fig:perf_comparision}
\end{figure*}

At FP16 precision, latency for all models are vastly improved, the fastest being \ModelName (121.9 FPS) and HarDNet-MSEG (118.7 FPS). In addition, accuracy metrics for all models are within $0.01\%$ of their FP32 counterparts. This further shows that large neural networks do not require high float precision to remain effective.

INT8 precision gives the most speed gain out of the three modes, but at great expense in terms of accuracy. \ModelName runs at 155.6 FPS in this mode, whereas HarDNet-MSEG is the second fastest model at 143.7 FPS. However, HarDNet-MSEG also suffers the largest drop in accuracy at about $8.1\%$ on all metrics. This drop is a bit lower for PraNet ($\approx 3.3\%$), NeoUNet ($\approx 6\%$) and \ModelName ($\approx 3.6\%$).

In every precision mode, \ModelName is consistently the fastest model while being a close second in terms of accuracy behind NeoUNet. \ModelName also displays its robustness to compression techniques, incurring significantly less degradation compared to models such as PraNet. Its small size also gives the proposed model advantage in long-term deployment, as energy usage and equipment wear become factors. In Figure~\ref{fig:perf_comparision}, we visualize the performance of BlazeNeo in terms of speed (FPS) and accuracy (dice score on polyp segmentation task) compared to other existing methods.

It should be emphasized that high FPS is essential in real scenarios because it helps endoscopists operate smoother and easier to monitor and detect lesions. According to endoscopists' experience, the idea FPS for colonoscopy should be at least 60. In addition, the fast inference speed gives us the potential to deploy the model on even more low-cost devices with less computational power, such as NVIDIA Jetson TX1/TX2, while still satisfying the minimal required FPS. This is important because it allows us to deploy the application on a large scale to many medical facilities in remote areas where the economic conditions are difficult.

%\section{Conclusion}
%\label{sec:conclude}
%This paper has proposed \ModelName, a novel neural network architecture for the polyp segmentation and neoplasm detection problem, with an emphasis on speed and deployability. \ModelName is an extremely lightweight and fast neural network, thanks to the use of an efficient HarDNet backbone, a multi-level feature aggregation structure, and an auxiliary training module to take advantage of undefined labels. Our experiments show that \ModelName outperforms all other state-of-the-art models in terms of inference latency while providing competitive accuracy. We also show that \ModelName can be very robust against degradation caused by compression techniques. In general, the proposed model is highly suitable for lightweight deployments with real-time requirements.

%\section{Appendixes}

For further discussion, we compare the performance of the models quantitatively and qualitatively in different precisions.
%\subsection{Quantitative Comparison of different precisions}

Table~\ref{tab:compress_acc} shows the accuracy of our \DHA models in different precisions. One can observe that the models with FP32 and FP16 precisions give the same result, which is lower than the original model by a mean margin of 2.32\%. The TensorRT INT8 model yields the worst accuracy, which is lower than the original model by a mean margin of 5.75\%.

\begin{table*}[ht!]
    \centering
    \caption{Accuracy metrics on the \CleanDatasetName test set for \DHA models in different precisions}
    \label{tab:compress_acc}
    \begin{tabular}{@{} c | r r r r r r r@{}}
        \toprule
        
        \multicolumn{1}{c|}{Model} & \multicolumn{1}{c}{$\text{Dice}_{seg}$} & \multicolumn{1}{c}{$\text{IoU}_{seg}$} & \multicolumn{1}{c}{$\text{Dice}_{non}$} & \multicolumn{1}{c}{$\text{IoU}_{non}$} & \multicolumn{1}{c}{$\text{Dice}_{neo}$} & \multicolumn{1}{c}{$\text{IoU}_{neo}$} \\
        \midrule
        \midrule
        PyTorch (w/o compression)                   & 0.904                                   & 0.825                                  & 0.717                                   & 0.559                                  & 0.885                                   & 0.792                                  \\
        TensorRT FP32              & 0.906                                   & 0.828                                  & 0.721                                   & 0.563                                  & 0.887                                   & 0.796                                  \\
        TensorRT FP16              & 0.906                                   & 0.828                                  & 0.721                                   & 0.563                                  & 0.887                                   & 0.796                                  \\
        TensorRT INT8              & 0.870                                   & 0.770                                  & 0.678                                   & 0.513                                  & 0.857                                   & 0.750                                  \\
        
        \bottomrule
    \end{tabular}
\end{table*}

With lower precision, a model can reduce latency, throughput and increase power efficiency. To compare the speed performance of the models in different precisions, we present detailed benchmarking results of them for 100 iterations on the Jetson AGX Xavier with two metrics: host latency and GPU compute time. Host latency is measured as the end-to-end execution time from the CPU point of view, while GPU computing time is the actual working time for GPU calculation.
\begin{figure*}[ht!]
    \begin{center}
        \includegraphics[scale=.33]{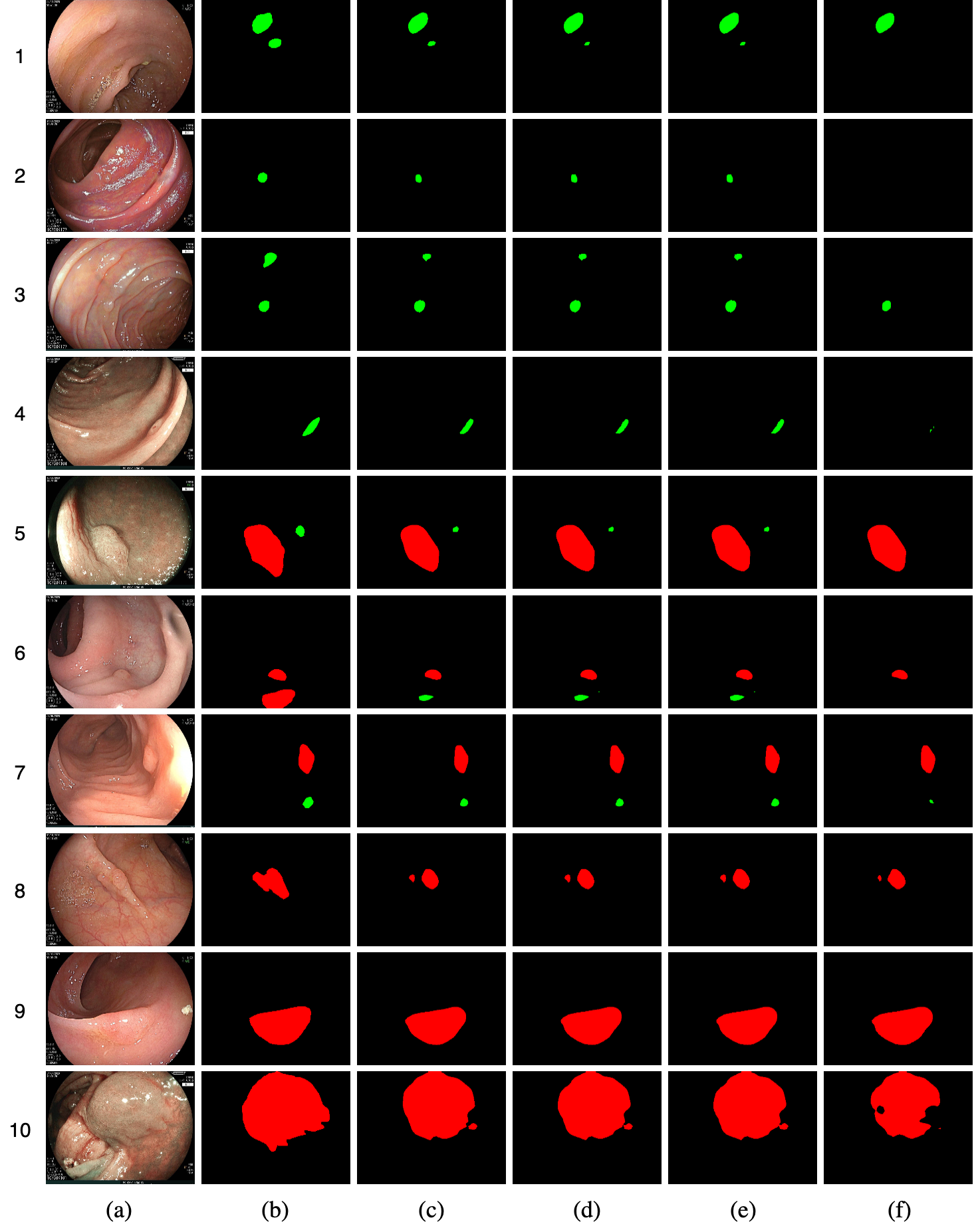}
    \end{center}
    \caption{Qualitative comparison of \DHA models in different precisions: (a) image, (b) ground truth, (c) Pytorch, (d) TensorRT FP32, (e) TensorRT FP16, (f) TensorRT INT8.}
    \label{fig:compress}
\end{figure*}

As shown in Table~\ref{tab:compress_latency}, the model in INT8 precision has the shortest latency compared to all other models, less than one-third of the model in FP32 precision. However, for deep learning inference, we must consider the trade-off between accuracy and speed. Therefore, the model in FP16 precision is the best choice, which gives the best result in all precisions and has a median latency.

\begin{table*}[ht!]
    \centering
    \caption{Latency metrics for \DHA models in different precisions.  The latency is measured on Jetson Xavier AGX with power mode MAXN}
    \label{tab:compress_latency}
    \begin{tabular}{l|c c c c|c c c c}
        \hline
        \multicolumn{1}{c|}{\multirow{2}{*}{Precision}} & \multicolumn{4}{c|}{\begin{tabular}[c]{@{}c@{}}Host Latency (ms)\end{tabular}} & \multicolumn{4}{c}{ \begin{tabular}[c]{@{}c@{}}GPU Compute (ms)\end{tabular}}                                                                                                 \\ \cline{2-9}
        \multicolumn{1}{c|}{}                           & min                                                 & max                                                 & mean          & median        & min           & max           & mean          & median        \\ \midrule
        \midrule
        TensorRT FP32                                   & 18.50                                               & 22.04                                               & 18.67         & 18.55         & 18.40         & 21.94         & 18.57         & 18.45         \\
        TensorRT FP16                                   & 8.10                                                & 10.24                                               & 8.20          & 8.15          & 7.99          & 10.13         & 8.09          & 8.04          \\
        TensorRT INT8                                   & \textbf{6.35}                                       & \textbf{6.51}                                       & \textbf{6.42} & \textbf{6.42} & \textbf{6.25} & \textbf{6.42} & \textbf{6.32} & \textbf{6.32} \\ \hline
    \end{tabular}
\end{table*}

%\subsection{Qualitative Comparison of different precisions}

For a more intuitive understanding of the loss in accuracy of each compressed model, Figure~\ref{fig:compress} illustrates the sample results of \DHA models in different precisions. As we can observe, the prediction result of models in FP32 and FP16 precisions are almost the same in every example. In general, the INT8 compressed model loses the ability to segment smaller polyps compared to other modes. For segmented polyps, the neoplasm prediction remains constant for every compression mode in all examples. This demonstrates that \ModelName can be quite robust to model compression and can be deployed in high compression modes such as FP16 with confidence.

\section{Conclusion}
\label{sec:conclude}
This paper has proposed \ModelName, a novel neural network architecture for the polyp segmentation and neoplasm detection problem, with an emphasis on speed and deployability. \ModelName is an extremely lightweight and fast neural network, thanks to the use of an efficient HarDNet backbone, a multi-level feature aggregation structure, and an auxiliary training module to take advantage of undefined labels. Our experiments show that \ModelName outperforms all other state-of-the-art models in terms of inference latency while providing competitive accuracy. We also show that \ModelName can be very robust against degradation caused by compression techniques. In general, the proposed model is highly suitable for lightweight deployments with real-time requirements. Our source code is available at \url{https://github.com/tofuai/neoplasm-segmentation}.

In future works, we would like to exploit recent advancements in Transformer-based architectures to improve the performance of the models. Especially, lightweight Transformers such as SegFormer \cite{xie2021segformer} will be the primary focus since they ensure both high accuracy and high inference speed which are crucial factors for a real computer-aided system in colonoscopy.  

\section{Acknowledgment}
This work was funded by Vingroup Innovation Foundation (VINIF) under project code VINIF.2020.DA17. %Phan Ngoc Lan was funded by Vingroup Joint Stock Company and supported by the Domestic Master/Ph.D. Scholarship Programme of Vingroup Innovation Foundation (VINIF), Vingroup Big Data Institute (VINBIGDATA), code VINIF.2020.ThS.BK.02.

%\clearpage

\bibliographystyle{splncs04}
\bibliography{paper}
%\balance
\begin{IEEEbiography}[{\includegraphics[width=1in,height=1.25in,clip,keepaspectratio]{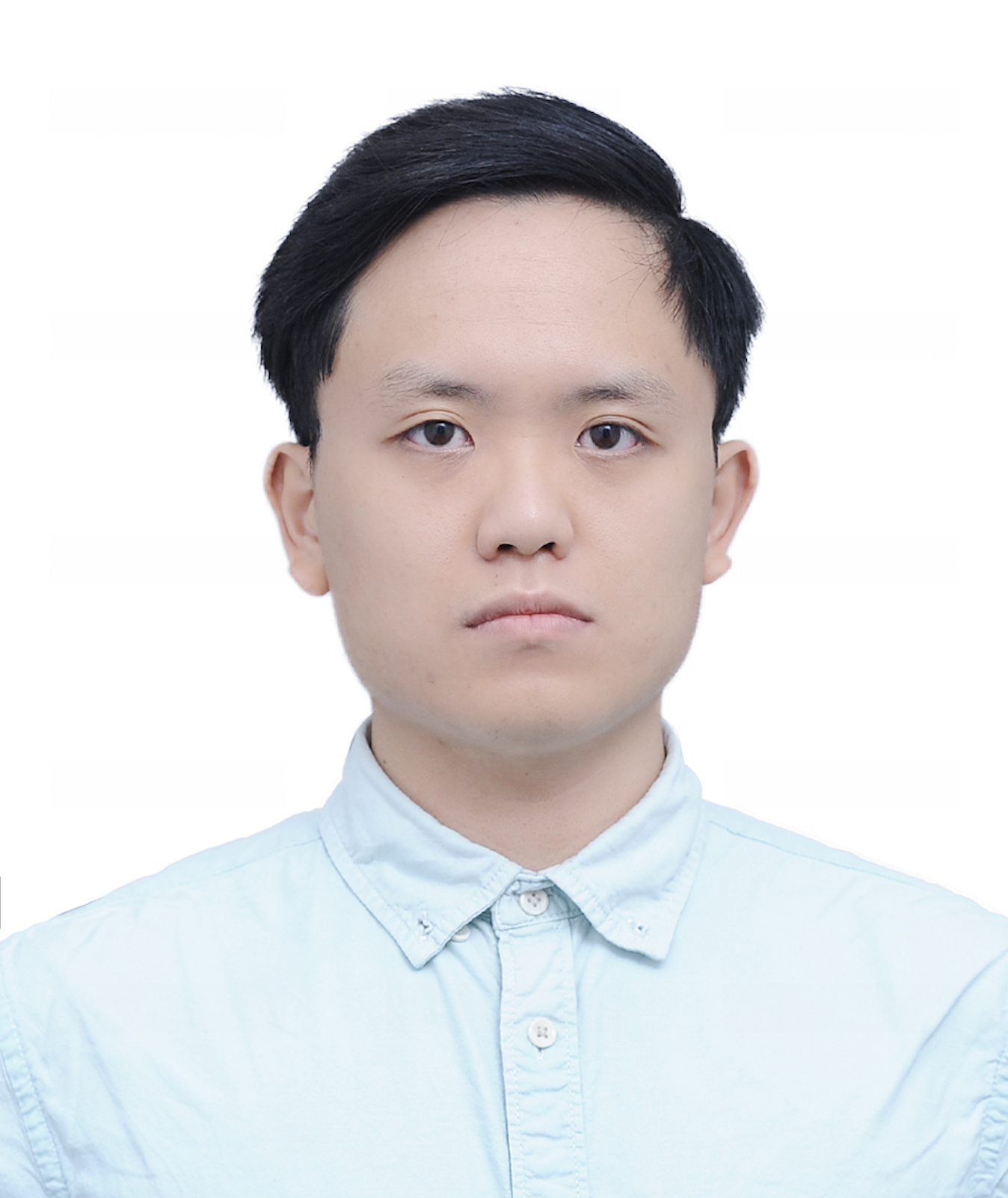}}]{Nguyen S. An} graduated the Global Engineering program in Information and Communication Technology, Hanoi University of Science and Technology (HUST). In 2018, he won the ERAMUS+ scholarship at Tampere University of Technology (TUT), Finland. Then, he was selected as one of the two Vietnamese students for participating in the Asia-Oceania Top University League (AOTULE) summer program 2019, at Bandung Institute of Technology.
\end{IEEEbiography}

\begin{IEEEbiography}[{\includegraphics[width=1in,height=1.25in,clip,keepaspectratio]{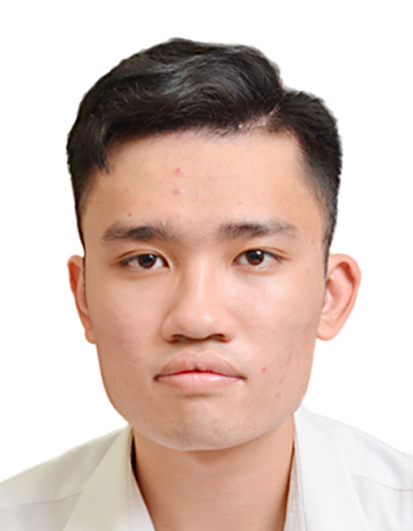}}]{Phan N. Lan} received his Engineering Degree in Information and Communication Technology at HUST in 2019. He is currently continuing his studies as a graduate student at HUST, majoring in Data Science and Artificial Intelligence. He is also a student member of the International Research Center for Artificial Intelligence (BK.AI). His research interests include machine learning, computer vision, and optimization techniques.
    
\end{IEEEbiography}

\begin{IEEEbiography}[{\includegraphics[width=1in,height=1.25in,clip,keepaspectratio]{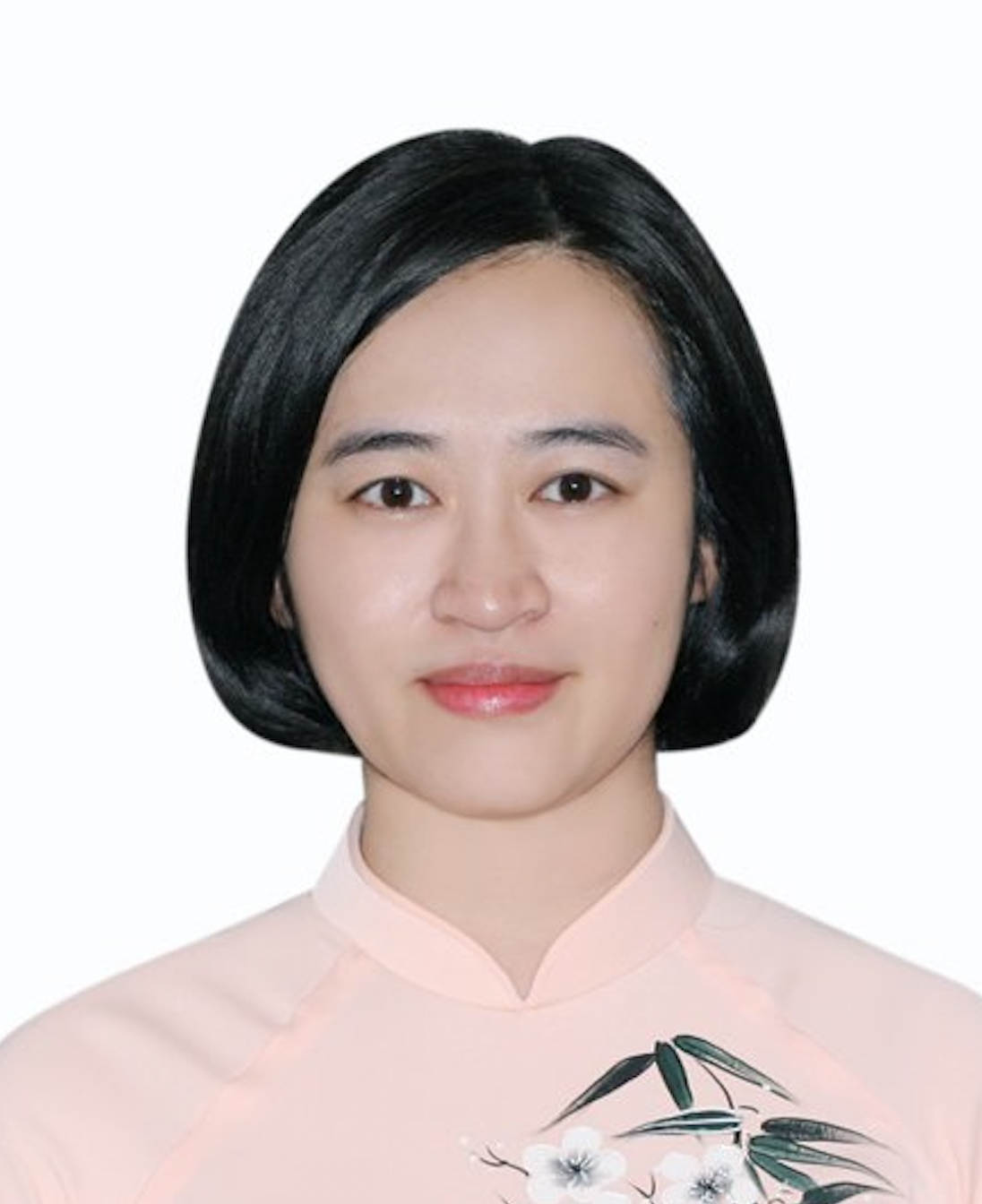}}]{Dao V. Hang} is a gastroenterologist and clinical lecturer at Hanoi Medical University and Institute of Gastroenterology and Hepatology. She also assumes different roles in the Vietnam Association of Gastroenterology (VNAGE), including organizing endoscopy training courses. Her current interest is the application of innovative technologies, such as image-enhanced endoscopy, smart apps, and artificial intelligence (AI) in endoscopy, microbiome, and GI motility. She believes technological solutions, among which AI is promising and feasible, can help solve the problems in limited-resources settings. She has experience participating in collaborative AI projects, where she led her team to recruit, label data, validate the product and implement clinical studies.
\end{IEEEbiography}

\begin{IEEEbiography}[{\includegraphics[width=1in,height=1.25in,clip,keepaspectratio]{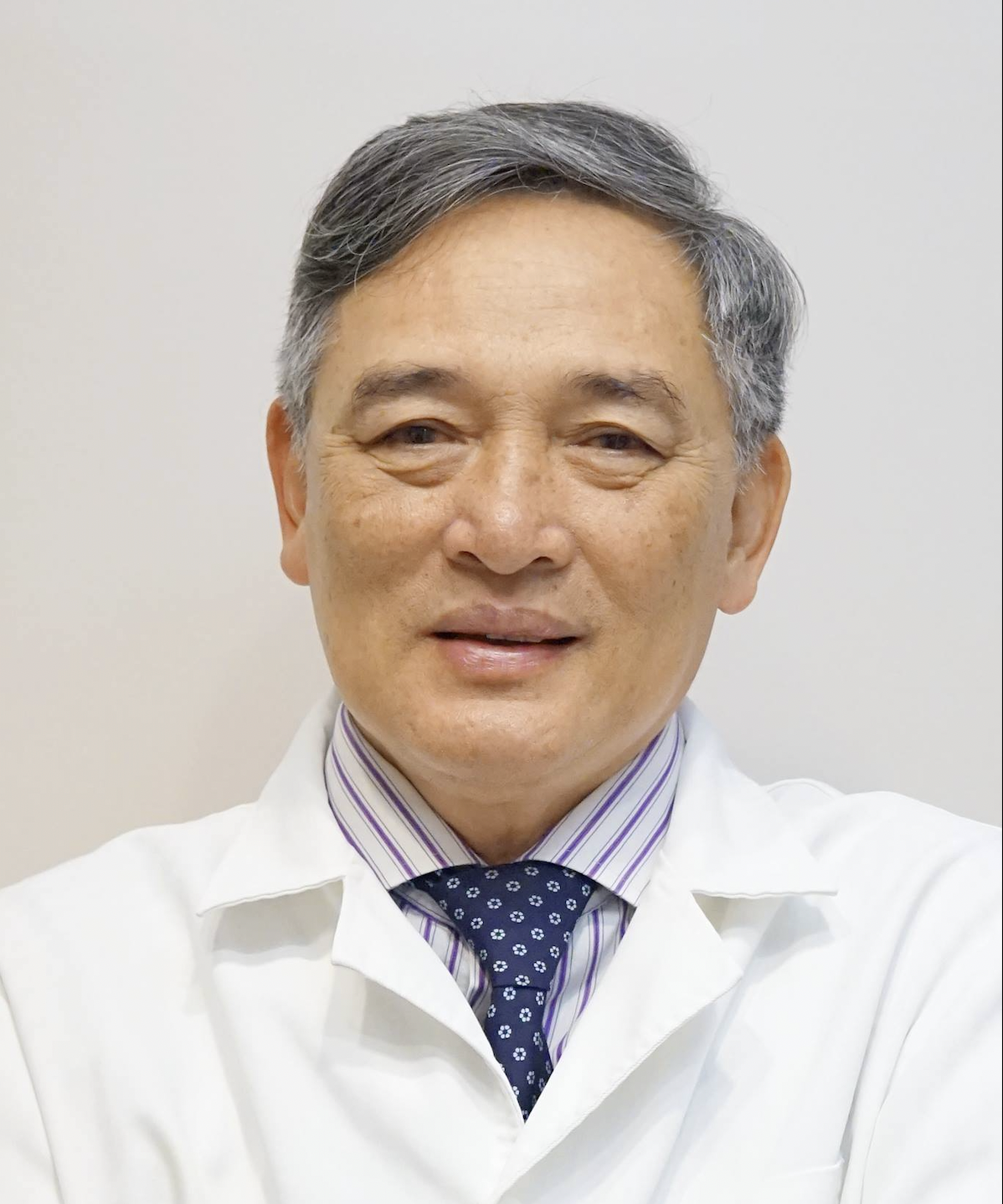}}]{Dao V. Long} is a professor in Gastroenterology and Hepatology and is now the Vice President of VNAGE. He was the Vice Rector of Hanoi Medical University, Director of Hanoi Medical University hospital, and Head of the Gastroenterology Department, Bach Mai hospital. He has good collaborations with foreign specialists and has organized many workshops and training courses for Vietnamese GI doctors and endoscopists. His research focuses on interventional endoscopy and GI tract diseases. During clinical practice, he has realized the importance of technology, particularly artificial intelligence, in GI endoscopy and has become interested in researching its application to facilitate diagnostic accuracy and reduce human labor in the practice of clinical physicians and endoscopists.
    
\end{IEEEbiography}

\begin{IEEEbiography}[{\includegraphics[width=1in,height=1.25in,clip,keepaspectratio]{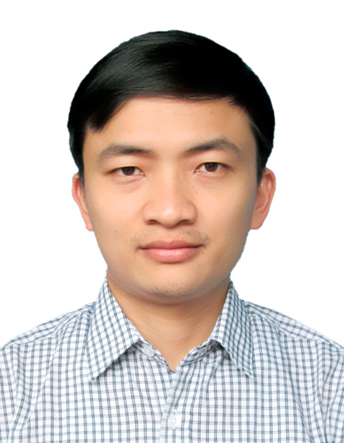}}]{Tran Q. Trung} is currently doing a research project in MD.PhD program at Greifswald University of Medicine, Germany. He is one of the young scientists who was selected globally to meet 70 Nobel Laureates at the 70th Lindau Nobel Meeting, where he can join a multidisciplinary network. He has been passionate about GI endoscopy since 2010, when he became a medical doctor. With experiences gained from advanced training in GI endoscopy in Vietnam, Japan, and Germany, and from a variety of international conferences in endoscopy, he is doing some novel and fruitful works for patients in Vietnam. The revolution of AI has recently attracted his great interest in applying to the field and ultimately bringing benefit to patients.
\end{IEEEbiography}

\begin{IEEEbiography}[{\includegraphics[width=1in,height=1.25in,clip,keepaspectratio]{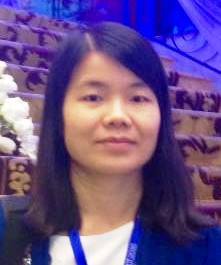}}]{Nguyen T. Thuy} received her Ph.D. degree in Computer Science from Graz University of Technology, Austria in 2009. She is now an Associate Professor, Head of Department of Computer Science, Faculty of Information Technology, Vietnam National University of Agriculture (VNUA), Vietnam. She has more than ten years of research experience in computer vision, machine learning, pattern recognition. She is an author and co-author of more than 70 research papers and patents. She has been a principal investigator and key member of a number of research projects in computer vision, machine learning, and applications.
\end{IEEEbiography}

\begin{IEEEbiography}[{\includegraphics[width=1in,height=1.25in,clip,keepaspectratio]{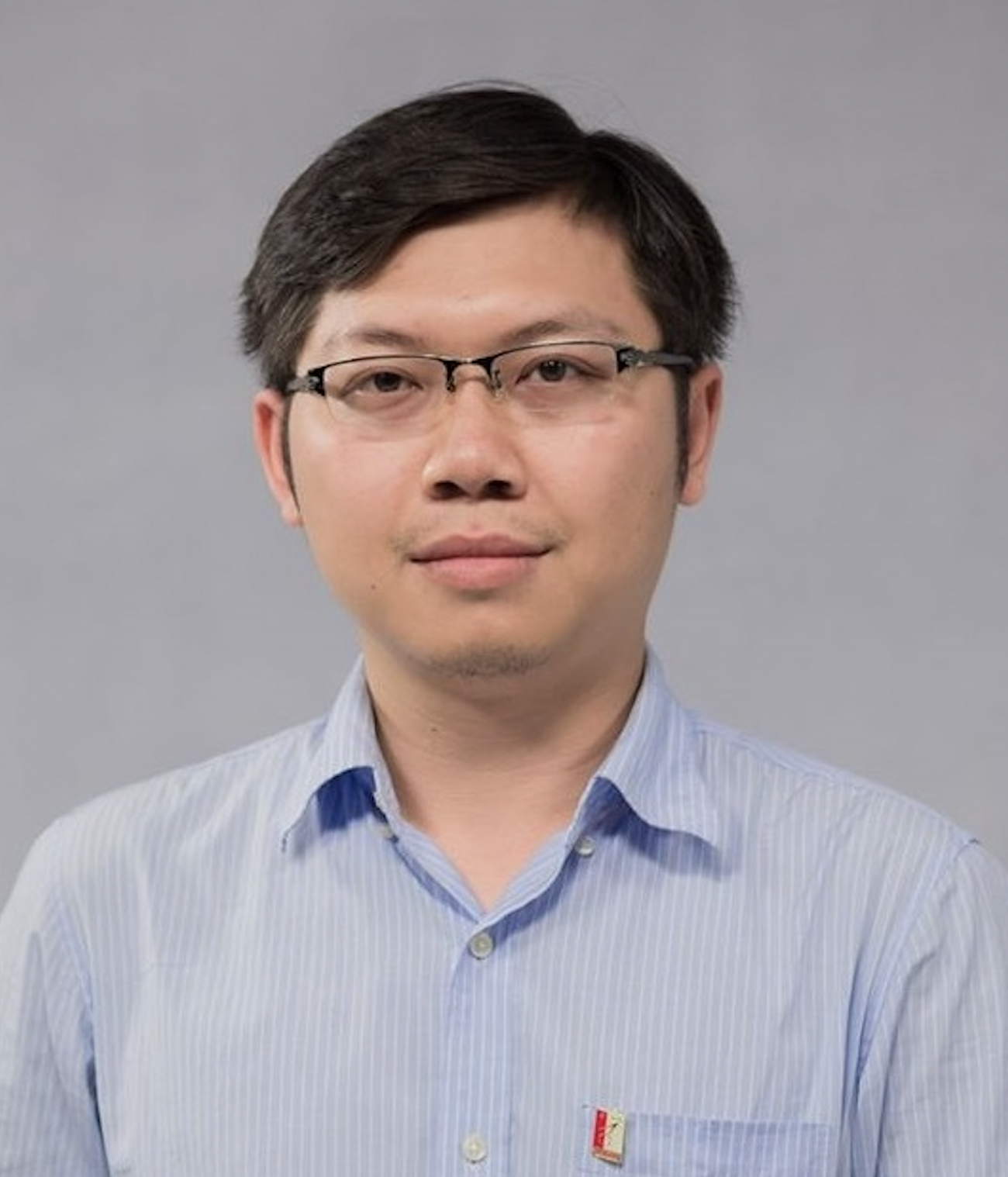}}]{Dinh V. Sang} received his Ph.D. degree in the Computer Science from Dorodnitsyn Computing Centre of the Russian Academy of Sciences (CCRAS) in 2013. He is currently working at the Department of Computer Science, School of Information and Communication Technology (SoICT), Hanoi University of Science and Technology (HUST), Vietnam. He is now the deputy managing director of the International Research Center for Artificial Intelligence (BK.AI) at HUST. His research interests include computer vision, machine learning, and deep learning. He has more than ten years of research experience in computer vision and machine learning and has published about 50 publications. He is the first NVIDIA deep learning institute (DLI) ambassador in Vietnam.
\end{IEEEbiography}

\EOD
\end{document}